\documentclass[reprint,amsmath,amssymb,prb]{revtex4-2}

\usepackage{graphicx}
\usepackage{xcolor}
\usepackage{tikz}
\usetikzlibrary{decorations.pathmorphing,decorations.markings,arrows.meta}

\usepackage{dcolumn}								
\usepackage{bm}
\usepackage{braket}
\usepackage{ulem}
\usepackage{amssymb,amsmath}
\usepackage{physics}
\usepackage{subfigure}
\usepackage{mathtools}
\usepackage{amsthm}

\usepackage{etoolbox}

\begin{document}
\title{Gauge invariance, collective modes, and the justification of normal-state subtraction in Dirac superconductors}

\author{Hiroshi Hayasaka} 
\email{h.hayasaka@rs.tus.ac.jp}
\affiliation{%
Department of Information Science and Technology, Tokyo University of Science, Noda, Chiba 278-8510, Japan\\
}%
\date{\today}
\begin{abstract}
In Dirac superconductors,
the unbounded spectrum of low-energy Dirac models is known to give rise to unphysical interband contributions from deep-lying states to the electromagnetic response.
To eliminate these contributions, normal-state subtraction (NSS), 
in which the normal-state response is subtracted from the superconducting-state response, has been widely employed. 
However, the relation between NSS and a gauge-invariant electromagnetic response, particularly the role of the vertex correction required by the Ward identity, has remained unclear.
In this work, we consider a massive-Dirac model with $s$-wave pairing and analytically investigate the electromagnetic response at zero temperature by solving the Bethe--Salpeter equation and incorporating the collective-mode contribution to the electromagnetic vertex.
We show that, in the static long-wavelength limit, the vertex correction exactly cancels the bare longitudinal response with NSS, yielding the vanishing longitudinal response required by gauge invariance.
In contrast, the transverse component of the vertex correction vanishes in the long-wavelength limit, so that the gauge-invariant Meissner weight coincides with that obtained from the bare transverse response with NSS.
Moreover, within the class of isotropic and analytic UV regularization terms, we show that gauge invariance uniquely fixes the regularization term in the static long-wavelength limit to the value prescribed by NSS.
Our results thus provide a microscopic justification for NSS.
\end{abstract}
\maketitle
\section{Introduction}
Dirac electron systems exhibit a variety of physical phenomena distinct from those in conventional electronic systems with parabolic dispersion, owing to their linear energy dispersion and characteristic band structure.
In Dirac superconductors, the interplay between the Dirac band structure and superconducting order is expected to give rise to unconventional electromagnetic responses.
The Meissner effect is a defining electromagnetic signature of superconductivity and has been extensively studied in Dirac superconductors~\cite{oudah2016superconductivity,leng2017type,teknowijoyo2018nodeless,guguchia2019nodeless,kopnin2008bcs}.

In conventional electronic systems with parabolic dispersion, both paramagnetic and diamagnetic currents contribute to the electromagnetic response.
In the normal state, these two contributions cancel each other, whereas in the superconducting state the cancellation is incomplete, resulting in a finite Meissner response.
By contrast, since a low-energy Dirac Hamiltonian is linear in momentum, it contains no quadratic coupling to the vector potential and lacks the usual diamagnetic term.
In the absence of the usual diamagnetic term, the Meissner response in Dirac electron systems is described by the paramagnetic current--current correlation.
In particular, interband effects are known to play an essential role in this response in Dirac systems~\cite{mizoguchi2015meissner}.

When the linearized Dirac spectrum is treated as unbounded, interband contributions from deep-lying states remain in the electromagnetic response and may produce a nonzero Meissner response even in the normal state.
Such contributions are considered artifacts of the low-energy approximation and are expected to be canceled in a theory that includes the full Brillouin zone and the high-energy band structure.
To avoid this problem, normal-state subtraction (NSS), in which the normal-state response is subtracted from the superconducting-state response, has been employed~\cite{uchoa2005nodal,kopnin2010supercurrent,mizoguchi2015meissner,schmidt2020odd}.

Uchoa et al.\ interpreted the subtraction of the normal-state response as a prescription that effectively restores the high-energy Brillouin-zone contributions missing in the cone approximation~\cite{uchoa2005nodal}.
In particular, they pointed out that the Meissner response in the normal state should vanish when the full Brillouin zone is taken into account and, based on the argument of Lifshitz and Pitaevskii~\cite{lifshitz2013statistical}, introduced a prescription in which the normal-state response is subtracted from the superconducting-state response.
From this viewpoint, NSS can be regarded as an ultraviolet (UV)
regularization prescription for the low-energy Dirac theory.

Mizoguchi and Ogata analyzed the Meissner response of three-dimensional massive-Dirac superconductors and showed that a finite Meissner kernel remains even in the normal state owing to the unbounded Dirac dispersion~\cite{mizoguchi2015meissner}.
They further introduced an extended model with a quadratic momentum term added to the Dirac Hamiltonian.
Using this model, they examined the validity of NSS by showing that the normal-state Meissner kernel vanishes and that the superconducting-state response agrees with the result obtained by NSS.

However, the bare current--current correlation is generally not sufficient to describe the electromagnetic response in a gauge-invariant manner.
In the superconducting state, collective modes involving phase fluctuations of the order parameter couple to the electromagnetic field, and the corresponding vertex correction must be included in the electromagnetic response.
Such a vertex correction is closely related to the Ward identity and plays an essential role in restoring gauge invariance, particularly in the longitudinal electromagnetic response~\cite{anderson1958coherent,bardasis1961excitons,nambu1960quasi}.

What remains unclear, however, is how NSS fits into a gauge-invariant electromagnetic response once the vertex correction is included.
More specifically, a conserving approximation ties the vertex correction to the mean-field self-energy~\cite{baym1961conservation}, whereas NSS is imposed directly on the bare response.
The key question is therefore whether, and how, NSS is compatible with the vertex correction required by gauge invariance.

In this work, we solve the Bethe--Salpeter equation (BSE) for a Dirac superconductor and analyze the electromagnetic response including the vertex correction associated with collective modes of the superconducting order parameter.
We show that, in the static long-wavelength limit at zero temperature, the vertex correction exactly cancels the bare longitudinal response with NSS, yielding a gauge-invariant electromagnetic response.
We also find that, 
in the same limit,
the transverse component of the vertex correction vanishes.
The same behavior was found for uniform $s$-wave superconductors in Ref.~\cite{boyack2020electromagnetic}.
Consequently, the Meissner weight obtained from the bare response using NSS coincides with that obtained from the gauge-invariant electromagnetic response including the vertex correction.
Furthermore, we show that, when the UV regularization term is isotropic and analytic near $\bm q=\bm 0$, gauge invariance uniquely fixes the regularization to the NSS prescription in the static long-wavelength limit.
This result provides a microscopic demonstration that combining the conventionally used NSS with the vertex correction yields a gauge-invariant electromagnetic response.

The remainder of this paper is organized as follows.
In Sec.~II, we discuss general aspects of the electromagnetic response in Dirac electron systems.
In Sec.~III, we introduce the model of a Dirac superconductor considered in this work.
In Sec.~IV, we define the electromagnetic response kernel within linear response theory.
In Sec.~V, we analyze the BSE for the electromagnetic vertex and present its solution together with the structure of the collective mode.
In Sec.~VI, we analyze the electromagnetic response including the vertex correction obtained from the BSE and clarify how gauge invariance is restored and how the response is related to NSS.
Finally, the appendices provide detailed derivations of several expressions used in the main text.
\section{General Framework for Gauge-Invariant Electromagnetic Response}
\label{sec:general-framework}

Before introducing a specific model, we outline the relation between gauge invariance, vertex corrections, and NSS in the electromagnetic response of superconductors.
Throughout this paper, the static limit is taken before the long-wavelength limit $q\to0$.

In an isotropic system, the linear response of the current to an external vector potential $A_{j}(\bm{q})$ is given by
\begin{align}
j_{i}(\bm{q})
&=
-\Phi_{ij}(\bm{q})A_{j}(\bm{q}),
\nonumber\\
\Phi_{ij}(\bm{q})
&=
\Phi_{\mathrm{T}}(\bm{q})
\left(
\delta_{ij}
-
\frac{q_{i}q_{j}}{q^{2}}
\right)
+
\Phi_{\mathrm{L}}(\bm{q})
\frac{q_{i}q_{j}}{q^{2}}.
\label{GeneralResponseDecomposition}
\end{align}
The quantities $\Phi_{\mathrm{T}}$ and $\Phi_{\mathrm{L}}$ denote the transverse and longitudinal responses, respectively.
Although a diamagnetic contribution is generally present for quadratic dispersions, the Dirac systems considered throughout this work have no conventional diamagnetic term, so that the bare electromagnetic response is given by the current--current correlation.
The transverse component describes the physical response to a magnetic field, and its long-wavelength limit gives the Meissner kernel.

For the physical current to remain unchanged under the gauge transformation
$A_{j}(\bm{q})\to A_{j}(\bm{q})+q_{j}\Lambda(\bm{q})$,
the response kernel must satisfy
\begin{align}
\Phi_{ij}(\bm{q})q_{j}
&=
0
\qquad\Longleftrightarrow\qquad
\Phi_{\mathrm{L}}(\bm{q})
=
0.
\label{GeneralGaugeCondition}
\end{align}
Therefore, the vanishing of the static longitudinal response provides a fundamental criterion for gauge invariance.

The electromagnetic response constructed solely from the mean-field Green function and the bare current vertex does not, in general, satisfy Eq.~(\ref{GeneralGaugeCondition}).
This is because introducing the mean-field self-energy while neglecting the corresponding vertex correction violates the Ward identity that should hold between the self-energy and the current vertex~\cite{nambu1960quasi,schrieffer2018theory}.
Constructing a gauge-invariant electromagnetic response requires the full vertex consistent with the mean-field self-energy so that the Ward identity is satisfied.

To separate the effects of UV regularization and the vertex correction, we write the full response as
\begin{align}
\Phi_{ij}^{\mathrm{GI}}(\bm{q})
&=
\Phi_{ij}^{(0),\mathrm{SC}}(\bm{q})
+
\Phi_{ij}^{\mathrm{UV}}(\bm{q})
+
\delta\Phi_{ij}^{\mathrm{VC}}(\bm{q}),
\label{GeneralPhysicalResponse}
\end{align}
where 
$\Phi_{ij}^{(0),\rm SC}$,
$\Phi_{ij}^{\mathrm{UV}}$, and
$\delta\Phi_{ij}^{\mathrm{VC}}$
denote the bare electromagnetic response kernel in the superconducting state, the UV regularization term, and the contribution of the vertex correction to the electromagnetic response kernel, respectively.
When the UV regularization preserves the Ward identity satisfied by the vertex, the full response satisfies Eq.~(\ref{GeneralGaugeCondition}), and its longitudinal component is required to satisfy
\begin{align}
\Phi_{\mathrm{L}}^{(0),\mathrm{SC}}(\bm{q})
+
\Phi_{\mathrm{L}}^{\mathrm{UV}}(\bm{q})
+
\delta\Phi_{\mathrm{L}}^{\mathrm{VC}}(\bm{q})
&=
0.
\label{GeneralLongitudinalCancellation}
\end{align}
Indeed, in continuum Dirac electron systems in the normal state, it has been pointed out that the choice of UV regularization can affect the gauge invariance of the electromagnetic response, and regularization schemes that preserve gauge invariance have also been proposed~\cite{fujimoto2013ultraviolet,takane2019gauge}.
In general, UV regularization schemes, such as those employed in Refs.~\cite{fujimoto2013ultraviolet}, need not take an additive form. In the present work, however, we assume that the UV regularization can be implemented as an additive contribution to the response kernel.

With NSS, the electromagnetic response kernel is defined as
\begin{align}
\Phi_{ij}^{\mathrm{sub}}(\bm{q})
&\equiv
\Phi_{ij}^{(0),\mathrm{SC}}(\bm{q})
-
\Phi_{ij}^{(0),\mathrm{N}}(\bm{q}),
\label{GeneralNSSDefinition}
\end{align}
where the superscript $\mathrm{N}$ denotes the normal state, namely, the kernel evaluated at vanishing superconducting gap $\Delta=0$.
Equation~(\ref{GeneralNSSDefinition}) corresponds to a prescription in which the bare normal-state response is adopted as the UV regularization term, i.e.,
$
\Phi_{ij}^{\mathrm{UV}}(\bm{q})
=
-\Phi_{ij}^{(0),\mathrm{N}}(\bm{q})
$
in Eq.~\eqref{GeneralPhysicalResponse}.
It should be noted that the Ward identity does not determine each individual contribution to the response.
In particular, the Ward identity alone does not single out the NSS prescription,
$
\Phi_{ij}^{\mathrm{UV}}(\bm{q})
=
-\Phi_{ij}^{(0),\mathrm{N}}(\bm{q}),
$
among possible UV regularizations.

When NSS is adopted as the
UV regularization scheme, Eq.~\eqref{GeneralLongitudinalCancellation} becomes
\begin{align}
\Phi_{\mathrm{L}}^{(0),\mathrm{SC}}(\bm{q})
-\Phi_{\mathrm{L}}^{(0),\mathrm{N}}(\bm{q})
+
\delta\Phi_{\mathrm{L}}^{\mathrm{VC}}(\bm{q})
&=
0.
\label{GeneralLongitudinalCancellation_sub}
\end{align}
We examine the relation between NSS and gauge invariance by checking whether Eq.~\eqref{GeneralLongitudinalCancellation_sub} is satisfied at $\omega=0$ and $q\to0$.
More specifically, by solving the BSE, we determine
$
\delta\Phi_{\mathrm{L}}^{\mathrm{VC}}(\bm{q})
$
and examine whether it exactly cancels
$
\Phi_{\mathrm{L}}^{(0),\mathrm{SC}}(\bm{q})
-\Phi_{\mathrm{L}}^{(0),\mathrm{N}}(\bm{q})
$.
Because the Ward identity does not constrain the transverse response, it remains to determine whether the vertex correction contains a transverse component.
As shown below, this can be established from the tensor structure of $\delta\Phi_{ij}^{\mathrm{VC}}$.

\section{Model}
In this section, following Mizoguchi and Ogata~\cite{mizoguchi2015meissner}, we summarize the formulation of the massive-Dirac superconducting model.
Most of the results presented in this section are standard or follow directly from Ref.~\cite{mizoguchi2015meissner}; they are included here to define the notation used throughout this paper and to make the subsequent derivations self-contained.
We consider the following three-dimensional massive-Dirac Hamiltonian~\cite{wolff1964matrix,fuseya2015transport}:
\begin{align}
\label{DiracHami}
{\cal H}_{0}
&=\left[
      \begin{array}{cc}
      M  & i\hbar v\bm{s} \cdot \bm{k}\\
      -i\hbar v\bm{s} \cdot \bm{k} &-M\\
      \end{array}
\right]
\nonumber\\
&=
M\rho_{3}\otimes s_{0}
-\hbar v\rho_{2}\otimes(\bm{k}\cdot\bm{s})
,
\end{align}
where $\bm{k}$, $v$, $2M$, and $\hbar$ denote the three-dimensional wave vector, a band parameter, the band gap, and the reduced Planck constant, respectively.
We assume $M>0$ and $v>0$.
Hereafter, we set $\hbar=1$ and the system volume to unity for simplicity.
The matrices $\bm{\rho}=(\rho_{1},\rho_{2},\rho_{3})$ and $\bm{s}=(s_{1},s_{2},s_{3})$ are Pauli matrices acting in the Bloch-band and spin spaces, respectively.
The basis in which the Hamiltonian is represented is 
$\{
\ket{1{\uparrow}}, 
\ket{1{\downarrow}}, 
\ket{2{\uparrow}}, 
\ket{2{\downarrow}}
\}
$, 
where $1$ and $2$ label the Bloch bands.
The symbols $\uparrow$ and $\downarrow$ denote the spin-up and spin-down states in the presence of spin--orbit coupling, respectively.
The unitary matrix that diagonalizes Eq.~(\ref{DiracHami}) is given by
\begin{align}
U(\bm{k})
&=
\begin{pmatrix}
Z_{\bm{k}}s_{0}
&
i\bm{Y}_{\bm{k}}\cdot\bm{s}
\\
i\bm{Y}_{\bm{k}}\cdot\bm{s}
&
Z_{\bm{k}}s_{0}
\end{pmatrix},
\end{align}
where
$
Z_{\bm{k}}
=
\sqrt{
(\epsilon_{\bm{k}}+M)
/(2\epsilon_{\bm{k}})
}
$
and
$
\bm{Y}_{\bm{k}}
=
v\bm{k}
/\sqrt{
2\epsilon_{\bm{k}}
\left(
\epsilon_{\bm{k}}+M
\right)
}
$.
The energy eigenvalues of Eq.~(\ref{DiracHami}) are given by
$\pm\epsilon_{\bm{k}}$,
where
$
\epsilon_{\bm{k}}
=
\sqrt{M^{2}+v^{2}k^{2}}
$.
Introducing the chemical potential $\mu$, we define
$
\tilde{h}_{0}(\bm{k})
=
{\cal H}_{0}
-\mu \rho_{0}\otimes s_{0}
$.
The second-quantized form of Eq.~(\ref{DiracHami}) is then
\begin{align}
H_{0}
&=
\sum_{\bm{k}}
\hat{c}_{\bm{k}}^{\dagger}
\tilde{h}_{0}(\bm{k})
\hat{c}_{\bm{k}}.
\end{align}
The fermionic annihilation operator is defined as
$
\hat{c}_{\bm{k}}
=
(
c_{\bm{k},1,\uparrow}, 
c_{\bm{k},1,\downarrow},
c_{\bm{k},2,\uparrow},
c_{\bm{k},2,\downarrow}
)^{\mathsf T}
$.
We also define the annihilation operator in the basis that diagonalizes Eq.~(\ref{DiracHami}) as
$
\hat{a}_{\bm{k}}
=
(
a_{\bm{k},+,\Uparrow},
a_{\bm{k},+,\Downarrow},
a_{\bm{k},-,\Uparrow},
a_{\bm{k},-,\Downarrow}
)
^{\mathsf T}
$.
Here, $\eta=+(-)$ labels the positive- (negative-) energy eigenstate of Eq.~(\ref{DiracHami}), while $\Uparrow$ and $\Downarrow$ denote the pseudospin associated with the doubly degenerate time-reversal and inversion partners.

For clarity, throughout this paper we refer to the bands with energies $\pm\epsilon_{\bm{k}}$ as the energy bands and to the corresponding eigenbasis as the energy-band basis.
By contrast, we refer to the basis in which Eq.~(\ref{DiracHami}) is represented as the Bloch-band basis.
In the energy-band basis,
\begin{align}
U(\bm{k})
\tilde{h}_{0}(\bm{k})
U^{\dagger}(\bm{k})
&=
\left(
\epsilon_{\bm{k}}\rho_{3}
-\mu\rho_{0}
\right)\otimes s_{0},
\end{align}
and the noninteracting Hamiltonian can be written as
\begin{align}
H_{0}
&=
\sum_{\bm{k}}
\sum_{\eta=\pm}
\sum_{\sigma=\Uparrow,\Downarrow}
\xi_{\eta}(\bm{k})
a_{\bm{k},\eta,\sigma}^{\dagger}
a_{\bm{k},\eta,\sigma},
\end{align}
where
$
\xi_{\eta}(\bm{k})
=
\eta\epsilon_{\bm{k}}-\mu
$.
Following Mizoguchi and Ogata, we assume an attractive $s$-wave interaction between time-reversal and inversion partners within the same energy band~\cite{mizoguchi2015meissner}.
We take the interaction Hamiltonian to be
\begin{align}
H_{\mathrm{int}}
&=
-V
\sum_{\bm{k},\bm{k}'}
f_{c}(k^{2})f_{c}(k'^{2})
\nonumber\\
&\times
\sum_{\eta,\eta'=\pm}
a_{\bm{k},\eta,\Uparrow}^{\dagger}
a_{-\bm{k},\eta,\Downarrow}^{\dagger}
a_{-\bm{k}',\eta',\Downarrow}
a_{\bm{k}',\eta',\Uparrow}.
\end{align}
Here, $V>0$ denotes the strength of the attractive interaction, and $f_{c}(k^{2})$ is a cutoff function that specifies the momentum range over which the interaction acts.
We adopt the following cutoff function:
\begin{align}
f_{c}(s)
&=
\frac{1}{2}
\left[
1-
\tanh\left(
\frac{s-k_{c}^{2}}
{\eta_{c}k_{c}^{2}}
\right)
\right].
\end{align}
This function approaches a step function in the limit $\eta_c\to 0^{+}$.
We define the superconducting order parameter in each energy band $\eta$ by
\begin{align}
\Delta_{\eta}
&=
V
\sum_{\bm{k}}
f_{c}(k^{2})
\left\langle
a_{-\bm{k},\eta,\Downarrow}
a_{\bm{k},\eta,\Uparrow}
\right\rangle,
\end{align}
and
\begin{align}
\Delta_{\bm{k}}
&=
\Delta f_{c}(k^{2}),
\end{align}
where
$
\Delta
=
\Delta_{+}
+
\Delta_{-}
$.
We choose the order parameter $\Delta$ to be real.
Applying the mean-field approximation to the full Hamiltonian $H_{0}+H_{\rm int}$, we obtain the following BCS Hamiltonian.
\begin{align}
H_{\mathrm{MF}}
&=
\sum_{\bm{k}}
\sum_{\eta=\pm}
\sum_{\sigma=\Uparrow,\Downarrow}
\xi_{\eta}(\bm{k})
a_{\bm{k},\eta,\sigma}^{\dagger}
a_{\bm{k},\eta,\sigma}
\nonumber\\
&\quad
-
\sum_{\bm{k}}
\sum_{\eta=\pm}
\Delta_{\bm{k}}
\left(
a_{\bm{k},\eta,\Uparrow}^{\dagger}
a_{-\bm{k},\eta,\Downarrow}^{\dagger}
+
a_{-\bm{k},\eta,\Downarrow}
a_{\bm{k},\eta,\Uparrow}
\right).
\end{align}
We define the Nambu spinors in the energy-band basis and the Bloch-band basis, respectively, as
\begin{align}
\Psi_{\bm{k}}
&=
\begin{pmatrix}
\hat{a}_{\bm{k}}\\
\hat{a}_{-\bm{k}}^{\dagger,\mathsf{T}}
\end{pmatrix},
&
\tilde{\Psi}_{\bm{k}}
&=
\begin{pmatrix}
\hat{c}_{\bm{k}}\\
\hat{c}_{-\bm{k}}^{\dagger,\mathsf{T}}
\end{pmatrix}.
\end{align}
The mean-field Hamiltonian can then be written, up to a constant term, as
\begin{align}
H_{\mathrm{MF}}
&=
\frac{1}{2}
\sum_{\bm{k}}
\Psi_{\bm{k}}^{\dagger}
\mathcal{H}_{\mathrm{BdG}}(\bm{k})
\Psi_{\bm{k}}.
\end{align}
The BdG Hamiltonian in the energy-band basis is
\begin{align}
\mathcal{H}_{\mathrm{BdG}}(\bm{k})
&=
\begin{pmatrix}
h_{0}(\bm{k})
&
-\Delta_{\bm{k}}\rho_{0}\otimes is_{2}
\\
-\Delta_{\bm{k}}
\left(
\rho_{0}\otimes is_{2}
\right)^{\dagger}
&
-h_{0}^{\mathsf{T}}(-\bm{k})
\end{pmatrix}
,
\end{align}
where 
$
h_0(\bm{k})
=
(\epsilon_{\bm{k}}\rho_3-\mu\rho_0)\otimes s_0
$.
The eigenenergies of the BdG Hamiltonian are
$
E_{\eta}(\bm{k})
=
\sqrt{
\xi_{\eta}^{2}(\bm{k})
+
\Delta_{\bm{k}}^{2}
}
$.
The self-consistency condition gives the gap equation
\begin{align}
1
&=
\frac{V}{2}
\sum_{\bm{k}}
f_{c}^{2}(k^{2})
\sum_{\eta=\pm}
\frac{1}{E_{\eta}(\bm{k})}
\tanh
\left(
\frac{E_{\eta}(\bm{k})}{2T}
\right).
\label{gapEq}
\end{align}
At $T=0$, this reduces to
\begin{align}
1
&=
\frac{V}{2}
\sum_{\bm{k}}
f_{c}^{2}(k^{2})
\left[
\frac{1}{E_{+}(\bm{k})}
+
\frac{1}{E_{-}(\bm{k})}
\right].
\label{gapEqZeroTemperature}
\end{align}
\section{Electromagnetic Response Function}
The bare velocity operator is given by
$
\tilde{j}_{i}
=
\partial \tilde{h}_{0}(\bm{k})/
\partial k_{i}
=
-v\rho_{2}\otimes s_{i}
$.
We define the corresponding bare current vertex in Nambu space as
\begin{align}
\tilde{\Gamma}_{i}^{(0)}
&=
\frac{1}{2}
\begin{pmatrix}
\tilde{j}_{i} & 0\\
0 &
-\tilde{j}_{i}^{\mathsf{T}}
\end{pmatrix}
.
\label{BareCurrentVertex}
\end{align}

\begin{figure}[t]
\centering
\resizebox{\columnwidth}{!}{%
\begin{tikzpicture}[
    line cap=round,
   fermion/.style={
  thick,
  postaction={decorate},
  decoration={
    markings,
    mark=at position 0.55 with {\arrow{Stealth[length=2.4mm]}}
  }
},
    photon/.style={thick,decorate,
       decoration={snake,amplitude=0.8mm,segment length=2.5mm}},
    pair/.style={thick,dashed},
    dot/.style={circle,fill=black,inner sep=1.55pt},
    every node/.style={font=\small}
]

\node[anchor=west] at (0,8.7) {(a)};
\begin{scope}[xshift=0.8cm]
\coordinate (vl) at (2.2,6.75);
\coordinate (vr) at (5.9,6.75);

\draw[
  thick,
  postaction={decorate},
  decoration={
    markings,
    mark=at position 0.53 with
    {\arrow{Stealth[length=3.6mm,width=2.4mm]}}
  }
] (vl)
 .. controls (2.3,8.1) and (5.8,8.1) .. (vr);

\draw[
  thick,
  postaction={decorate},
  decoration={
    markings,
    mark=at position 0.54 with
    {\arrow{Stealth[length=3.6mm,width=2.4mm]}}
  }
] (vr)
 .. controls (5.8,5.4) and (2.3,5.4) .. (vl);

\draw[photon] (0.85,6.75) -- (vl);
\draw[photon] (vr) -- (7.25,6.75);

\node[dot] at (vl) {};

\node[
  thick,
  draw,
  circle,
  fill=white,
  minimum size=10mm,
  inner sep=0pt
] at (vr) {$\tilde{\Gamma}_j$};

\node[above] at (1.12,6.9) {$i$};
\node[above] at (6.95,6.9) {$j$};
\node[above] at (4.05,7.85) {$\tilde{\mathcal{G}}_+$};
\node[below] at (4.05,5.65) {$\tilde{\mathcal{G}}_-$};
\node at (1.85,6.23) {$\tilde{\Gamma}_i^{(0)}$};
\end{scope}

\node[anchor=west] at (0,4.77) {(b)};
\begin{scope}[yshift=0.9cm]
\draw[thick] (0.1,0.65) -- (1.05,1.05);
\draw[thick] (1.05,1.05) -- (2.0,0.65);
\draw[photon] (1.05,1.05) -- (1.05,2.65);
\node[
  draw,
  thick,
  circle,
  fill=white,
  minimum size=10mm,
  inner sep=0pt
] at (1.05,1.05) {$\tilde{\Gamma}_i$};

\node at (2.4,1.05) {$=$};

\draw[thick] (2.8,0.65) -- (3.75,1.05);
\draw[thick] (3.75,1.05) -- (4.7,0.65);
\node[dot] at (3.75,1.05) {};
\draw[photon] (3.75,1.05) -- (3.75,2.65);
\node[below] at (3.75,0.9) {$\tilde{\Gamma}_i^{(0)}$};

\node at (5.22,1.05) {$-$};
\node at (5.93,1.05) {$\displaystyle\sum_{\sigma=\pm}$};
\end{scope}

\draw[thick] (7.05,0.65) -- (8.10,1.05);
\draw[thick] (8.10,1.05) -- (9.15,0.65);
\node[dot] at (8.10,1.05) {};
\node[below] at (8.10,0.90) {$V_{\sigma}$};
\node[below] at (8.10,0.90) {$V_{\sigma}$};

\draw[pair] (8.10,1.05) -- (8.10,2.02);
\node[right] at (8.10,1.52) {$V$};

\coordinate (bottom) at (8.10,2.02);
\coordinate (top) at (8.10,3.40);

\draw[
  thick,
  postaction={decorate},
  decoration={
    markings,
    mark=at position 0.5 with {
      \pgftransformxshift{1.2mm}
      \arrow{Stealth[length=2.4mm]}
    }
  }
] (bottom)
 .. controls (7.15,2.14) and (7.15,3.28) .. (top);

\draw[
  thick,
  postaction={decorate},
  decoration={
    markings,
    mark=at position 0.5 with {
      \pgftransformxshift{1.2mm}
      \arrow{Stealth[length=2.4mm]}
    }
  }
] (top)
 .. controls (9.05,3.28) and (9.05,2.14) .. (bottom);

\node[dot] at (bottom) {};
\node[left] at (7.25,2.7) {$\tilde{\mathcal{G}}_+$};
\node[right] at (8.93,2.7) {$\tilde{\mathcal{G}}_-$};
\node[above right,xshift=-4mm,yshift=1mm] at (bottom)
 {$V_{-\sigma}$};

\draw[photon] (top) -- (8.10,4.45);

\node[
  draw,
  thick,
  circle,
  fill=white,
  minimum size=10mm,
  inner sep=0pt
] at (top) {$\tilde{\Gamma}_i$};

\end{tikzpicture}%
}
\caption{
Diagrammatic representation of (a) the current--current response
and (b) the Bethe--Salpeter equation for the dressed current vertex.
The upper and lower propagators in (a) are the full Nambu Green functions
$
\tilde{\mathcal G}_+
\equiv
\tilde{\mathcal G}(\bm k_+,i\epsilon_n)
$
and
$
\tilde{\mathcal G}_-
\equiv
\tilde{\mathcal G}(\bm k_-,i\epsilon_{n-}),
$
 respectively. 
Wavy lines denote external
field insertions. In (b), the two pairing vertices are
$V_+=\tau_+\otimes B(\boldsymbol k)$ and
$V_-=\tau_-\otimes B^\dagger(\boldsymbol k)$,
and the dashed line denotes the separable interaction
$V f_c(k^2)f_c(k'^{2})$. The fermion loop includes a trace and the
momentum--frequency sum, as in Eqs.~(27) and (28).
Panel (b) represents the BSE in the static limit
$i\omega_\lambda=0$.
}
\label{fig:vertex-bse}
\end{figure}
The current--current correlation function is given by the Kubo formula,
\begin{align}
&
\Phi_{ij}(\bm{q},i\omega_{\lambda})
=
2e^{2}T
\sum_{\bm{k},n}
\mathrm{Tr}_{N,\rho,s}
\Bigl[
\tilde{\mathcal{G}}
(\bm{k}_{-},i\epsilon_{n-})
\tilde{\Gamma}_{i}^{(0)}
\nonumber\\
&\qquad\qquad\times
\tilde{\mathcal{G}}
(\bm{k}_{+},i\epsilon_{n})
\tilde{\Gamma}_{j}
(\bm{k}_{+},\bm{k}_{-};
i\epsilon_{n},i\epsilon_{n-})
\Bigr],
\label{Kuboformula}
\end{align}
where $e>0$ is the elementary charge, so that the electron charge is $-e$.
We have defined
$
\bm{k}_{\pm}
=
\bm{k}\pm\bm{q}/2
$
and
$
i\epsilon_{n-}
=
i\epsilon_{n}-i\omega_{\lambda}
$,
where $i\omega_{\lambda}$ is a bosonic Matsubara frequency.
The current--current correlation function is
diagrammatically illustrated in Fig.~\ref{fig:vertex-bse}(a).
The transformation between the energy-band and Bloch-band Nambu bases is implemented by
\begin{align}
\mathcal{U}_{\rm BdG}(\bm{k})
&=
\begin{pmatrix}
U(\bm{k}) & 0\\
0 & U^{*}(-\bm{k})
\end{pmatrix}
.
\label{UnitaryBdG}
\end{align}
The Green function in the Bloch-band basis is given by
$\tilde{\mathcal{G}}
=
\mathcal{U}^{\dagger}_{\rm BdG}
\mathcal{G}
\mathcal{U}_{\rm BdG}$,
where
$\mathcal{G}(\bm{k},i\epsilon_n)
=
[i\epsilon_n-\mathcal{H}_{\mathrm{BdG}}(\bm{k})]^{-1}$
is the Green function in the energy-band basis.
$
\tilde{\Gamma}_{j}
$
is the vertex function, namely, the current vertex including the vertex correction.
\section{Bethe--Salpeter Equation}
In this section, we solve the BSE and obtain an explicit expression for the vertex function and its components.
We consider the static limit $i\omega_{\lambda}=0$.
In this limit, the two fermionic frequencies entering the vertex are equal.
We display only a single fermionic frequency for the vertex function and write
$
\tilde{\Gamma}_{i}
(\bm{k}_{+},\bm{k}_{-};i\epsilon_n,i\epsilon_n)
=
\tilde{\Gamma}_{i}
(\bm{k}_{+},\bm{k}_{-};i\epsilon_n)
$.
Since the external momentum $\bm q=\bm k_+-\bm k_-$ is uniquely determined by $\bm k_\pm$, we do not display it as an independent argument.
The BSE corresponding to the model considered in this work is given by
\begin{align}
&
\tilde{\Gamma}_{i} (\bm{k}_{+},\bm{k}_{-};i\epsilon_{n})
=
\tilde{\Gamma}^{(0)}_{i}(\bm{k}_{+},\bm{k}_{-};i\epsilon_{n})
\nonumber\\
&\quad
-
f_{c}(k^{2})
\left[
\tau_{+}\otimes B(\bm{k})\,C_{i}^{+}(\bm{q})
+
\tau_{-}\otimes B^{\dagger}(\bm{k})\,C_{i}^{-}(\bm{q})
\right],
\label{BSEBeforePauliExpansion}
\end{align}
where
$
\tau_{\pm}
=
(\tau_{1}\pm i\tau_{2})/2
$
are the raising and lowering operators in Nambu space, $\bm{\tau}=(\tau_{1},\tau_{2},\tau_{3})$ denotes the Pauli matrices acting in Nambu space and 
$B(\bm{k})$ is given by
\begin{align}
B(\bm{k})
&=
\frac{M}{\epsilon_{\bm{k}}}
\rho_{0}\otimes is_{2}
-
\frac{iv}{\epsilon_{\bm{k}}}
\rho_{1}\otimes
(\bm{k}\cdot\bm{s})is_{2}.
\end{align}
The quantities $C^{+}$ and $C^{-}$ are defined as
\begin{align}
C_{i}^{+}(\bm{q})
&=
\frac{VT}{2}
\sum_{\bm{k}',m}
f_{c}(k'^{2})
e^{-i\epsilon_{m}0^{-}}
\mathrm{Tr}_{\rho,s}
\Biggl[
B^{\dagger}(\bm{k}')
\nonumber\\
&\times
\Bigl[
\tilde{\mathcal{G}}(\bm{k}'_+,i\epsilon_{m})
\tilde{\Gamma}_{i}
(\bm{k}_{+}',\bm{k}_{-}';i\epsilon_{m})
\tilde{\mathcal{G}}(\bm{k}'_-,i\epsilon_{m})
\Bigr]_{12}
\Biggr],
\label{CPlusDefinition}
\\
C_{i}^{-}(\bm{q})
&=
\frac{VT}{2}
\sum_{\bm{k}',m}
f_{c}(k'^{2})
e^{-i\epsilon_{m}0^{-}}
\mathrm{Tr}_{\rho,s}
\Biggl[
B(\bm{k}')
\nonumber\\
&\times
\Bigl[
\tilde{\mathcal{G}}(\bm{k}'_+,i\epsilon_{m})
\tilde{\Gamma}_{i}
(\bm{k}_{+}',\bm{k}_{-}';i\epsilon_{m})
\tilde{\mathcal{G}}(\bm{k}'_-,i\epsilon_{m})
\Bigr]_{21}
\Biggr].
\label{CMinusDefinition}
\end{align}
The symbol $\mathrm{Tr}_{\rho,s}$ denotes the trace over the Bloch-band and spin spaces.
The corresponding BSE is diagrammatically illustrated
in Fig.~\ref{fig:vertex-bse}(b).
This BSE is obtained by functionally differentiating the Dyson equation with respect to the external field~\cite{baym1961conservation}.
This construction ensures the required consistency between the mean-field self-energy and the vertex correction.
For details of the derivation of Eqs.~(\ref{BSEBeforePauliExpansion})--(\ref{CMinusDefinition}), see Appendix~\ref{apdxA}.

We now recast the BSE as a closed $2\times2$ system for $C_i^+(\bm q)$ and $C_i^-(\bm q)$.
For convenience, we define
$
V_+(\bm k)=\tau_+\otimes B(\bm k)
$
and
$
V_-(\bm k)=\tau_-\otimes B^\dagger(\bm k)
$,
and introduce the shorthand
$
\tilde{\mathcal G}_\pm
\equiv
\tilde{\mathcal G}(\bm k_\pm,i\epsilon_n)
$.
Then, Eq.~\eqref{BSEBeforePauliExpansion} can be written as
\begin{align}
\tilde{\Gamma}_i
&=
\tilde{\Gamma}_i^{(0)}
-
f_c(k^2)
\left[
V_+C_i^+(\bm q)
+
V_-C_i^-(\bm q)
\right].
\label{BSEDirectCompact}
\end{align}
Substituting Eq.~\eqref{BSEDirectCompact} into Eq.~\eqref{CPlusDefinition}, we obtain
\begin{align}
&
C_i^+(\bm q)
\nonumber\\
&=
\frac{VT}{2}
\sum_{\bm k,n}
f_c(k^2)e^{-i\epsilon_n0^-}
\operatorname{Tr}_{N,\rho,s}
\left[
\tilde{\mathcal G}_-
V_-
\tilde{\mathcal G}_+
\tilde{\Gamma}_i^{(0)}
\right]
\nonumber\\
&-
\frac{VT}{2}
\sum_{\bm k,n}
f_c^2(k^2)e^{-i\epsilon_n0^-}
\operatorname{Tr}_{N,\rho,s}
\left[
\tilde{\mathcal G}_-
V_-
\tilde{\mathcal G}_+
V_+
\right]
C_i^+(\bm q)
\nonumber\\
&-
\frac{VT}{2}
\sum_{\bm k,n}
f_c^2(k^2)e^{-i\epsilon_n0^-}
\operatorname{Tr}_{N,\rho,s}
\left[
\tilde{\mathcal G}_-
V_-
\tilde{\mathcal G}_+
V_-
\right]
C_i^-(\bm q).
\label{DirectCPlusEquation}
\end{align}
Similarly, substituting Eq.~\eqref{BSEDirectCompact} into Eq.~\eqref{CMinusDefinition}, we obtain
\begin{align}
&
C_i^-(\bm q)
\nonumber\\
&
=
\frac{VT}{2}
\sum_{\bm k,n}
f_c(k^2)e^{-i\epsilon_n0^-}
\operatorname{Tr}_{N,\rho,s}
\left[
\tilde{\mathcal G}_-
V_+
\tilde{\mathcal G}_+
\tilde{\Gamma}_i^{(0)}
\right]
\nonumber\\
&-
\frac{VT}{2}
\sum_{\bm k,n}
f_c^2(k^2)e^{-i\epsilon_n0^-}
\operatorname{Tr}_{N,\rho,s}
\left[
\tilde{\mathcal G}_-
V_+
\tilde{\mathcal G}_+
V_+
\right]
C_i^+(\bm q)
\nonumber\\
&-
\frac{VT}{2}
\sum_{\bm k,n}
f_c^2(k^2)e^{-i\epsilon_n0^-}
\operatorname{Tr}_{N,\rho,s}
\left[
\tilde{\mathcal G}_-
V_+
\tilde{\mathcal G}_+
V_-
\right]
C_i^-(\bm q).
\label{DirectCMinusEquation}
\end{align}
We define the source terms generated by the bare vertex as
\begin{align}
&
\chi_i^+(\bm q)
\nonumber\\
&
\equiv
\frac{VT}{2}
\sum_{\bm k,n}
f_c(k^2)e^{-i\epsilon_n0^-}
\operatorname{Tr}_{N,\rho,s}
\left[
\tilde{\mathcal G}_-
V_-
\tilde{\mathcal G}_+
\tilde{\Gamma}_i^{(0)}
\right],
\nonumber\\
&
\chi_i^-(\bm q)
\nonumber\\
&\equiv
\frac{VT}{2}
\sum_{\bm k,n}
f_c(k^2)e^{-i\epsilon_n0^-}
\operatorname{Tr}_{N,\rho,s}
\left[
\tilde{\mathcal G}_-
V_+
\tilde{\mathcal G}_+
\tilde{\Gamma}_i^{(0)}
\right]
\label{SourceTermDefinition}
\end{align}
and the kernels that couple $C_i^+$ and $C_i^-$ as
\begin{align}
&
\Pi^{++}(\bm q)
\nonumber\\
&\equiv
-\frac{VT}{2}
\sum_{\bm k,n}
f_c^2(k^2)e^{-i\epsilon_n0^-}
\operatorname{Tr}_{N,\rho,s}
\left[
\tilde{\mathcal G}_-
V_-
\tilde{\mathcal G}_+
V_+
\right],
\nonumber\\
&
\Pi^{+-}(\bm q)
\nonumber\\
&\equiv
-\frac{VT}{2}
\sum_{\bm k,n}
f_c^2(k^2)e^{-i\epsilon_n0^-}
\operatorname{Tr}_{N,\rho,s}
\left[
\tilde{\mathcal G}_-
V_-
\tilde{\mathcal G}_+
V_-
\right],
\nonumber\\
&
\Pi^{-+}(\bm q)
\nonumber\\
&\equiv
-\frac{VT}{2}
\sum_{\bm k,n}
f_c^2(k^2)e^{-i\epsilon_n0^-}
\operatorname{Tr}_{N,\rho,s}
\left[
\tilde{\mathcal G}_-
V_+
\tilde{\mathcal G}_+
V_+
\right],
\nonumber\\
&
\Pi^{--}(\bm q)
\nonumber\\
&
\equiv
-\frac{VT}{2}
\sum_{\bm k,n}
f_c^2(k^2)e^{-i\epsilon_n0^-}
\operatorname{Tr}_{N,\rho,s}
\left[
\tilde{\mathcal G}_-
V_+
\tilde{\mathcal G}_+
V_-
\right].
\label{BSEKernelDefinition}
\end{align}
The coupled equations can then be collected into the following $2\times2$ matrix equation:
\begin{align}
\begin{pmatrix}
1-\Pi^{++}(\bm q)
&
-\Pi^{+-}(\bm q)
\\
-\Pi^{-+}(\bm q)
&
1-\Pi^{--}(\bm q)
\end{pmatrix}
\begin{pmatrix}
C_i^+(\bm q)
\\
C_i^-(\bm q)
\end{pmatrix}
&=
\begin{pmatrix}
\chi_i^+(\bm q)
\\
\chi_i^-(\bm q)
\end{pmatrix}.
\label{ClosedBSEMatrix}
\end{align}
Symmetry further reduces this matrix equation.
For the present model,
$
\mathcal{S}
=
\tau_0
\otimes
\rho_0
\otimes
is_2
$ 
satisfies 
$
\mathcal{S}
\tilde{\mathcal G}^{\mathsf T}(\bm{k},i\epsilon_n)
\mathcal{S}^{-1}
=
\tilde{\mathcal G}(\bm{k},i\epsilon_n),
$
$
\mathcal{S} 
V_-^{\mathsf T}(\bm{k})
\mathcal{S}^{-1}
=
V_+(\bm{k}).
$
Using these relations, we find
\begin{align}
\operatorname{Tr}
\left[
\tilde{\mathcal G}_{-}V_{-}
\tilde{\mathcal G}_{+}V_{-}
\right]
&=
\operatorname{Tr}
\left[
\tilde{\mathcal G}_{-}V_{+}
\tilde{\mathcal G}_{+}V_{+}
\right],
\end{align}
and hence
$
\Pi^{+-}(\bm q)=\Pi^{-+}(\bm q)
$. 
For the inversion operator
$
\mathcal{I}=
\tau_0
\otimes
\rho_3
\otimes s_0
$,
the relations
$
\tilde{\mathcal G}(-\bm{k},i\epsilon_n)
=
\mathcal{I}
\tilde{\mathcal G}(\bm{k},i\epsilon_n)\mathcal{I}^{-1}
$
and
$
V_{\pm}(-\bm{k})
=
\mathcal{I} V_{\pm}(\bm{k})\mathcal{ I}^{-1}
$
hold.
Using the change of variables $\bm{k}\to-\bm{k}$ and the cyclic property of the trace, we obtain
\begin{align}
\Pi^{++}(\bm q)
&=
-\frac{VT}{2}
\sum_{\bm{k},n}
f_c^2(k^2)e^{-i\epsilon_n0^-}
\operatorname{Tr}
\left[
\tilde{\mathcal G}_{+}V_{-}
\tilde{\mathcal G}_{-}V_{+}
\right]
\nonumber\\
&
=
\Pi^{--}(\bm q).
\end{align}
It is therefore natural to rotate to the symmetric and antisymmetric combinations,
\begin{align}
C_{s,i}(\bm{q})
=
\frac{
C_{i}^{+}(\bm{q})
+
C_{i}^{-}(\bm{q})
}{\sqrt{2}},
\quad
C_{a,i}(\bm{q})
=
\frac{
C_{i}^{+}(\bm{q})
-
C_{i}^{-}(\bm{q})
}{\sqrt{2}},
\end{align}
\begin{align}
\chi_{s,i}(\bm{q})
=
\frac{
\chi_{i}^{+}(\bm{q})
+
\chi_{i}^{-}(\bm{q})
}{\sqrt{2}},
\quad
\chi_{a,i}(\bm{q})
=
\frac{
\chi_{i}^{+}(\bm{q})
-
\chi_{i}^{-}(\bm{q})
}{\sqrt{2}},
\end{align}
\begin{align}
&\Pi_{s}(\bm{q})
=
\Pi^{++}(\bm{q})+\Pi^{+-}(\bm{q}),
\end{align}
\begin{align}
&\Pi_{a}(\bm{q})
=
\Pi^{++}(\bm{q})-\Pi^{+-}(\bm{q}).
\label{AntisymmetricComponents}
\end{align}
In this basis, 
the BSE decouples into
\begin{align}
\left[
1-\Pi_{a}(\bm{q})
\right]
C_{a,i}(\bm{q})
&=
\chi_{a,i}(\bm{q})
\label{AntisymmetricBSE}
\end{align}
and
\begin{align}
\left[
1-\Pi_{s}(\bm{q})
\right]
C_{s,i}(\bm{q})
&=
\chi_{s,i}(\bm{q})
.
\label{SymmetricBSE}
\end{align}
Using the gap equation~(\ref{gapEqZeroTemperature}), we find that, at $\bm{q}=0$, the antisymmetric channel satisfies
$
1-\Pi_{a}(\bm{0})
=
0
$.
This means that the collective mode is gapless at $\bm{q}=0$.
For $\chi_{a,i}$, the long-wavelength expansion gives the following result:
\begin{align}
\chi_{a,i}(\bm{q})
&=
-\frac{Vv^{2}\Delta}{12\sqrt{2}}
S_{1}q_{i}
+
O(q^{3}),
\label{eq:chi_a_explicit}
\end{align}
where
$S_{1}=\sum_{\bm{k}}\mathcal{S}_{1}(\bm{k})$,
with $\mathcal{S}_{1}(\bm{k})$ given by
\begin{align}
\mathcal{S}_{1}(\bm{k})
&=
\Biggl[
\frac{
v^{2}k^{2}f_{c}^{2}(k^{2})
}{
\epsilon_{\bm{k}}^{2}
}
\left(
\frac{1}{E_{-}^{3}(\bm{k})}
+
\frac{1}{E_{+}^{3}(\bm{k})}
\right)
\nonumber\\
&
+
\frac{
2k^{2}f_{c}(k^{2})
}{
\epsilon_{\bm{k}}
}
\left.
\frac{df_{c}(s)}{ds}
\right|_{s=k^{2}}
\left(
\frac{\xi_{-}(\bm{k})}{E_{-}^{3}(\bm{k})}
-
\frac{\xi_{+}(\bm{k})}{E_{+}^{3}(\bm{k})}
\right)
\Biggr]
.
\label{eq:def_S1}
\end{align}
The following relation also holds for the ratio of $\chi_{a,i}$ to $1-\Pi_a(\bm{q})$:
\begin{align}
\frac{\chi_{a,i}(\bm{q})}
{1-\Pi_{a}(\bm{q})}
&=
-\sqrt{2}\Delta
\frac{q_{i}}{q^{2}}
+
O(q).
\label{eq:ratio_long_wavelength}
\end{align}
The derivation of Eqs.~(\ref{eq:chi_a_explicit})
and
(\ref{eq:ratio_long_wavelength}) is given in Appendix~\ref{apdxB}.
Using Eq.~\eqref{eq:ratio_long_wavelength}, the solution of Eq.~\eqref{AntisymmetricBSE} takes the form
\begin{align}
C_{a,i}(\bm{q})
&=
-
\sqrt{2}\Delta
\frac{q_{i}}{q^{2}}
+
O(q).
\label{AntisymmetricVertexCoefficient}
\end{align}
For the symmetric channel, $1-\Pi_s(\bm{0})\neq0$, and the corresponding correction vanishes as $q\to0$, as shown in Appendix~\ref{apdxC}; only the singular antisymmetric contribution proportional to $C_{a,i}$ survives.

We define the pairing vertex corresponding to the antisymmetric channel as
\begin{align}
\tilde{V}_{a}(\bm{k})
&=
\tau_{+}\otimes B(\bm{k})
-
\tau_{-}\otimes B^{\dagger}(\bm{k}).
\label{AntisymmetricPairVertex}
\end{align}
The vertex function can be written, apart from regular vertex corrections that vanish in the long-wavelength limit, as
\begin{align}
&
\tilde{\Gamma}_{i}
(\bm{k}_{+},\bm{k}_{-};i\epsilon_{n})
=
\tilde{\Gamma}_{i}^{(0)}
+
\delta\tilde{\Gamma}_{i}
(\bm{k}_{+},\bm{k}_{-};i\epsilon_{n})
,
\label{FullCurrentVertex}
\end{align}
where
\begin{align}
\delta\tilde{\Gamma}_{i}
(\bm{k}_{+},\bm{k}_{-};i\epsilon_{n})
=
\Delta
f_{c}(k^{2})
\frac{q_{i}}{q^{2}}
\tilde{V}_{a}(\bm{k}).
\label{CurrentVertexCorrection}
\end{align}
Thus, in addition to the bare current vertex, the full vertex contains a singular collective-mode correction.

\section{Gauge-Invariant Electromagnetic Response and Normal-State Subtraction}
\label{GaugeInvariance}
In this section, we show how the vertex correction combines with NSS to yield a gauge-invariant electromagnetic response.
We then examine whether gauge invariance uniquely fixes the UV regularization term.
We write the static response kernel as $\Phi_{ij}(\bm{q})=\Phi_{ij}(\bm{q},0)$.
We also omit all regular contributions that vanish as $q\to0$, including the contribution from the symmetric channel discussed in Appendix~\ref{apdxC}.

Substituting Eq.~(\ref{FullCurrentVertex}) into the Kubo formula~(\ref{Kuboformula}) 
gives
\begin{align}
\Phi_{ij}^{\mathrm{SC}}(\bm{q})
&=
\Phi_{ij}^{(0),\mathrm{SC}}(\bm{q})
+
\delta\Phi_{ij}^{\mathrm{VC}}(\bm{q}).
\label{ResponseWithVertexDecomposition}
\end{align}
The contribution of the vertex correction is given by
\begin{align}
&
\delta\Phi_{ij}^{\mathrm{VC}}(\bm{q})
\nonumber\\
&=
2e^{2}T
\sum_{\bm{k},n}
\mathrm{Tr}_{N,\rho,s}
\Bigl[
\tilde{\mathcal{G}}
(\bm{k}_{-},i\epsilon_{n})
\tilde{\Gamma}_{i}^{(0)}
\tilde{\mathcal{G}}
(\bm{k}_{+},i\epsilon_{n})
\nonumber\\
&\hspace{42mm}\times
\delta\tilde{\Gamma}_{j}
(\bm{k}_{+},\bm{k}_{-};i\epsilon_{n})
\Bigr].
\label{BSECurrentCorrectionDefinition}
\end{align}
Substituting Eq.~(\ref{CurrentVertexCorrection}) into Eq.~(\ref{BSECurrentCorrectionDefinition}), we obtain
\begin{align}
&
\delta\Phi_{ij}^{\mathrm{VC}}(\bm{q})
\nonumber\\
&=
e^{2}\Delta
\frac{q_{j}}{q^{2}}
\Biggl[
2T
\sum_{\bm{k},n}
f_{c}(k^{2})
\mathrm{Tr}_{N,\rho,s}
\Bigl[
\tilde{\mathcal{G}}
(\bm{k}_{-},i\epsilon_{n})
\tilde{\Gamma}_{i}^{(0)}
\nonumber\\
&\quad\times
\tilde{\mathcal{G}}
(\bm{k}_{+},i\epsilon_{n})
\tilde{V}_{a}(\bm{k})
\Bigr]
\Biggr].
\label{BSECorrectionMixedCorrelation}
\end{align}
Equation~(\ref{BSECorrectionMixedCorrelation}) can be expressed in terms of $\chi_{a,i}$, yielding
\begin{align}
\delta\Phi_{ij}^{\mathrm{VC}}(\bm{q})
&=
-
e^{2}\Delta
\frac{q_{j}}{q^{2}}
\left[
\frac{
4\sqrt{2}}{V}
\chi_{a,i}(-\bm{q})
\right]
\nonumber\\
&=
-\frac{e^{2}\Delta^{2}v^{2}}{3}
S_{1}
\frac{q_{i}q_{j}}{q^{2}}.
\label{BSELongitudinalCorrection}
\end{align}
Since this contribution is proportional to $q_{i}q_{j}/q^{2}$, it is purely longitudinal.
Thus,
\begin{align}
\delta\Phi_{\mathrm{L}}^{\mathrm{VC}}(\bm{q})
&=
-\frac{e^{2}\Delta^{2}v^{2}}{3}
S_{1}.
\label{BSELongitudinalCorrection_2}
\end{align}
For the NSS introduced in Eq.~(\ref{GeneralNSSDefinition}), evaluating the longitudinal component of the bare electromagnetic response kernel in the present model gives
\begin{align}
\Phi_{\mathrm{L}}^{\mathrm{sub}}(\bm{q})
&=
\frac{
e^{2}\Delta^{2}v^{2}
}{3}
S_{1}.
\label{SubtractedLongitudinalResponseExplicit}
\end{align}
This result is obtained from the bare response kernels in the superconducting and normal states derived by Mizoguchi and Ogata~\cite{mizoguchi2015meissner}.
The details of the calculation are given in Appendix~\ref{apdxD}.

Combining Eqs.~\eqref{BSELongitudinalCorrection_2} and \eqref{SubtractedLongitudinalResponseExplicit}, the longitudinal component in the long-wavelength limit becomes
\begin{align}
\lim_{\bm{q}\to0}
\Phi_{\mathrm{L}}^{\mathrm{GI}}(\bm{q})
&=
\lim_{\bm{q}\to0}
[
\Phi_{\mathrm{L}}^{\mathrm{sub}}(\bm{q})
+
\delta\Phi_{\mathrm{L}}^{\mathrm{VC}}(\bm{q})
]
\nonumber\\
&=
0.
\label{LongitudinalWardCancellation}
\end{align}
The exact cancellation between the vertex correction and the bare longitudinal response with NSS yields the vanishing longitudinal electromagnetic response required by gauge invariance.

The tensor structure of Eq.~(\ref{BSELongitudinalCorrection}) also determines the transverse response.
Since its leading term is proportional to $q_{i}q_{j}/q^{2}$, its projection onto the transverse direction, $(\delta_{ij}-q_{i}q_{j}/q^2)$, satisfies
\begin{align}
\lim_{q\to 0}
\left(
\delta_{i\ell}
-
\frac{q_{i}q_{\ell}}{q^{2}}
\right)
\delta\Phi_{\ell j}^{\mathrm{VC}}(\bm{q})
&=
0.
\label{TransverseBSEProjection}
\end{align}
Equation~(\ref{TransverseBSEProjection}) shows that the singular vertex correction does not contribute to the transverse Meissner kernel.
Thus,
\begin{align}
\lim_{q\to 0}
\delta\Phi_{\mathrm{T}}^{\mathrm{VC}}(\bm{q})
&=
0.
\label{TransverseBSE}
\end{align}
This vanishing transverse contribution agrees with the general result for uniform $s$-wave superconductors reported in Ref.~\cite{boyack2020electromagnetic}.

The vanishing of the transverse vertex correction then implies that the gauge-invariant transverse response coincides with the transverse response obtained by NSS, i.e.,
\begin{align}
\lim_{q\to0}
\Phi_{\mathrm{T}}^{\mathrm{GI}}(\bm{q})
&=
\lim_{q\to0}
\Phi_{\mathrm{T}}^{\mathrm{sub}}(\bm{q}).
\label{MeissnerKernelNSS}
\end{align}
This equality shows that the Meissner kernel with NSS is the transverse component of the gauge-invariant electromagnetic response including the vertex correction.

Finally, we discuss the uniqueness of the UV regularization term.
From the results obtained above, its longitudinal component is uniquely fixed as
$
\Phi_{\mathrm{L}}^{\mathrm{UV}}
=
-\Phi_{\mathrm{L}}^{(0),\mathrm{N}}
$
as $q\to 0$
.
If this term is isotropic and analytic in the vicinity of $\bm q=\bm 0$, its longitudinal and transverse components have the same limit:
\begin{align}
\lim_{q\to0}
\Phi_{\mathrm{T}}^{\mathrm{UV}}(\bm q)
&=
\lim_{q\to0}
\Phi_{\mathrm{L}}^{\mathrm{UV}}(\bm q)
\nonumber\\
&=
-\lim_{q\to0}
\Phi_{\mathrm{L}}^{(0),\mathrm{N}}(\bm q).
\end{align}
Since the normal-state response is also isotropic and analytic near $\bm q=\bm 0$, we have 
$
\lim_{q\to0}\Phi_{\mathrm{L}}^{(0),\mathrm{N}}
=
\lim_{q\to0}\Phi_{\mathrm{T}}^{(0),\mathrm{N}}
$.
It follows that
\begin{align}
\lim_{q\to0}
\Phi_{\mathrm{T}}^{\mathrm{UV}}(\bm q)
&=
-\lim_{q\to0}
\Phi_{\mathrm{T}}^{(0),\mathrm{N}}(\bm q).
\end{align}
Thus, the transverse component is also uniquely fixed and coincides with that prescribed by NSS.
Consequently, under these conditions, the Meissner response is uniquely fixed to the value prescribed by NSS.
\section{Conclusion}
In low-energy Dirac theories, the unbounded spectrum produces unphysical interband contributions that remain in the response.
Conventionally, these contributions have been removed using NSS.

In this work, we analyzed a massive-Dirac superconductor with $s$-wave pairing at zero temperature.
We solved the BSE derived from the superconducting self-energy and examined the collective-mode contribution to the electromagnetic vertex.

We showed that, in the static long-wavelength limit, the vertex correction exactly cancels the bare longitudinal response with NSS, yielding a gauge-invariant electromagnetic response.
The transverse component of the vertex correction vanishes, so that the response evaluated with NSS and the gauge-invariant electromagnetic response including the vertex correction yield the same Meissner weight.
For an isotropic and analytic UV regularization term, gauge invariance uniquely fixes the regularization to the NSS prescription, and hence uniquely determines the corresponding Meissner response.
These results provide a microscopic justification for applying NSS to a massive-Dirac superconductor with $s$-wave pairing.

Although we have focused on the static long-wavelength limit, 
an important question is whether and how the relation between NSS and the vertex-corrected electromagnetic response persists at finite wave vector and frequency. 
A further question is how broadly the relation established here extends to more general multiband superconductors with different pairing symmetries and to other Dirac and Weyl systems. 
Clarifying these issues would further establish the scope of NSS as an effective description of gauge-invariant electromagnetic response in multiband superconductors.
\appendix
\section{Derivation of the Bethe--Salpeter equation}
\label{apdxA}
In this Appendix, we derive the BSE in the Bloch-band basis.
The BSE can be obtained by functionally differentiating the Dyson equation~\cite{baym1961conservation}.
We first derive the representation of $\mathcal{H}_{\rm BdG}$ in the Bloch-band basis and the corresponding self-energy.
The Nambu spinors in the energy-band basis and the Bloch-band basis are related through Eq.~\eqref{UnitaryBdG} as
$
\Psi_{\bm{k}}
=
\mathcal{U}_{\mathrm{BdG}}(\bm{k})
\tilde{\Psi}_{\bm{k}}
$.

We define the matrix obtained by transforming the spin-singlet pairing matrix from the energy-band basis to the Bloch-band basis as
\begin{align}
B(\bm{k})
&=
U^{\dagger}(\bm{k})
\left(
\rho_{0}\otimes is_{2}
\right)
U^{\dagger,\mathsf{T}}(-\bm{k})
\nonumber\\
&=
\frac{M}{\epsilon_{\bm{k}}}
\rho_{0}\otimes is_{2}
-
\frac{iv}{\epsilon_{\bm{k}}}
\rho_{1}\otimes
(\bm{k}\cdot\bm{s})is_{2}.
\end{align}
Its Hermitian conjugate is
\begin{align}
B^{\dagger}(\bm{k})
&=
-\frac{M}{\epsilon_{\bm{k}}}
\rho_{0}\otimes is_{2}
-
\frac{iv}{\epsilon_{\bm{k}}}
\rho_{1}\otimes
is_{2}(\bm{k}\cdot\bm{s})
.
\label{PairingMatrixBdagger}
\end{align}
The BdG Hamiltonian in the Bloch-band basis can be written as
\begin{align}
\tilde{\mathcal{H}}_{\mathrm{BdG}}(\bm{k})
&=
\begin{pmatrix}
\tilde{h}_{0}(\bm{k})
&
-\Delta_{\bm{k}}B(\bm{k})
\\
-\Delta_{\bm{k}}B^{\dagger}(\bm{k})
&
-\tilde{h}_{0}^{\mathsf{T}}(-\bm{k})
\end{pmatrix}.
\label{OrbitalBdGHamiltonian}
\end{align}

We next allow the order parameter to vary spatially in the presence of an external field.
We denote the external field by $\mathcal{A}_{i}$, whose relation to the vector potential $A_i$ used in the main text is given by
$
\mathcal{A}_{i}
\equiv
-2eA_i
$.

To describe spatially nonuniform pairing, we adopt the following finite-$\bm{q}$ extension of the interaction introduced in Sec. III: 
\begin{align}
H_{\mathrm{int}}
&=
-V\sum_{\bm{q}}P_{\bm{q}}^{\dagger}P_{\bm{q}},
\\
P_{\bm{q}}^{\dagger}
&=
\frac{1}{2}
\sum_{\bm{k}}
f_{c}(k^{2})
\hat{c}_{\bm{k}_{+}}^{\dagger}
B(\bm{k})
\hat{c}_{-\bm{k}_{-}}^{\dagger,\mathsf{T}},
\\
P_{\bm{q}}
&=
\frac{1}{2}
\sum_{\bm{k}}
f_{c}(k^{2})
\hat{c}_{-\bm{k}_{-}}^{\mathsf{T}}
B^{\dagger}(\bm{k})
\hat{c}_{\bm{k}_{+}}.
\end{align}
Here, $\bm{k}$ is the relative momentum of the two electrons forming a pair, while $\bm{q}$ is the total momentum of the pair.
In this extension, the internal form factor 
$f_{c}(k^2)B(\bm{k})$ is taken to depend only on the relative momentum $\bm{k}$, while the center-of-mass momentum $\bm{q}$ enters through the fermionic momenta.
We define the order parameter in the presence of the external field as
$
\Delta_{\bm{q}}^{\mathcal{A}}
\equiv
V
\left\langle
P_{\bm{q}}
\right\rangle_{\mathcal{A}}
$.
Corresponding to this finite-$\bm{q}$ pairing interaction and the definition of the order parameter, the BdG Hamiltonian in the Bloch-band basis in the presence of the external field is given by
\begin{align}
&
\tilde{\mathcal{H}}_{\mathrm{BdG}}^{\mathcal{A}}
(\bm{k}_{+},\bm{k}_{-})
\nonumber\\
&=
\begin{pmatrix}
\tilde{h}_{0}^{\mathcal{A}}(\bm{k}_{+},\bm{k}_{-}) &
-\Delta_{\bm{q}}^{\mathcal{A}}f_{c}(k^{2})B(\bm{k})
\\
-\Delta_{-\bm{q}}^{\mathcal{A}*}f_{c}(k^{2})B^{\dagger}(\bm{k}) &
-\tilde{h}_{0}^{\mathcal{A},\mathsf{T}}(-\bm{k}_{-},-\bm{k}_{+})
\end{pmatrix}.
\label{ExternalFieldOrbitalBdGHamiltonian}
\end{align}
We define the BdG Hamiltonian without the pairing term as
\begin{align}
&
\tilde{\mathcal{H}}_{0,\mathrm{BdG}}^{\mathcal{A}}
(\bm{k}_{+},\bm{k}_{-})
\nonumber\\
&=
\begin{pmatrix}
\tilde{h}_{0}^{\mathcal{A}}(\bm{k}_{+},\bm{k}_{-}) &
0
\\
0 &
-\tilde{h}_{0}^{\mathcal{A},\mathsf{T}}(-\bm{k}_{-},-\bm{k}_{+})
\end{pmatrix}
.
\label{ExternalFieldNormalOrbitalBdGHamiltonian}
\end{align}
The self-energy in the presence of the external field is therefore given by
\begin{align}
&
\tilde{\Sigma}^{\mathcal{A}}
(\bm{k}_{+},\bm{k}_{-})
\nonumber\\
&=
\tilde{\mathcal{H}}_{\mathrm{BdG}}^{\mathcal{A}}
(\bm{k}_{+},\bm{k}_{-})
-
\tilde{\mathcal{H}}_{0,\mathrm{BdG}}^{\mathcal{A}}
(\bm{k}_{+},\bm{k}_{-})
\nonumber\\
&=
-f_c(k^2)
\left[
\tau_{+}\otimes B(\bm{k})
\Delta_{\bm{q}}^{\mathcal{A}}
+
\tau_{-}\otimes B^{\dagger}(\bm{k})
\Delta_{-\bm{q}}^{\mathcal{A}*}
\right]
.
\label{OrbitalSelfEnergy}
\end{align}
For the BSE derivation, it is convenient to switch to four-momentum notation.
We write the external field as $\mathcal{A}_{i}(q)$ and define the external four-momentum as
$
q
=
(\bm{q},i\omega_{\lambda})
$.
We also define
$
k_{+}
=
(\bm{k}_{+},i\epsilon_{n})
$
and
$
k_{-}
=
(\bm{k}_{-},i\epsilon_{n}-i\omega_{\lambda})
$.
Since translational symmetry is broken in the presence of the external field, the Nambu Green function generally depends on two four-momenta and is written as
\begin{align}
\tilde{\mathcal{G}}^{\mathcal{A}}(k_{1},k_{2})
&=
-
\left\langle
T_{\tau}
\tilde{\Psi}_{k_{1}}
\tilde{\Psi}_{k_{2}}^{\dagger}
\right\rangle_{\mathcal{A}}.
\end{align}
In the four-momentum representation, we denote the two gap amplitudes corresponding to the 12 and 21 blocks by
$
\Delta_{12}^{\mathcal{A}}(q)
$
and
$
\Delta_{21}^{\mathcal{A}}(q)
$,
respectively.
They correspond to
$
\Delta_{\bm{q}}^{\mathcal{A}}
$
and
$
\Delta_{-\bm{q}}^{\mathcal{A}*}
$,
respectively, in the spatial-momentum representation, and Hermiticity requires
\begin{align}
\Delta_{21}^{\mathcal{A}}(q)
&=
\left[
\Delta_{12}^{\mathcal{A}}(-q)
\right]^{*}.
\end{align}
The four-momentum gap amplitudes corresponding to those appearing in Eq.~\eqref{OrbitalSelfEnergy} are determined self-consistently from the finite-$\bm{q}$ pairing interaction.
In terms of the anomalous components of the Nambu Green function, they can be written as
\begin{align}
\Delta_{12}^{\mathcal{A}}(q)
&=
\frac{VT}{2}
\sum_{\bm{k}',m}
f_{c}(k'^{2})
e^{-i\epsilon_{m}0^{-}}
\nonumber\\
&\quad\times
\mathrm{Tr}_{\rho,s}
\left[
B^{\dagger}(\bm{k}')
\left[
\tilde{\mathcal{G}}^{\mathcal{A}}
(k_{+}',k_{-}')
\right]_{12}
\right],
\label{GeneralizedGapPlus}
\\
\Delta_{21}^{\mathcal{A}}(q)
&=
\frac{VT}{2}
\sum_{\bm{k}',m}
f_{c}(k'^{2})
e^{-i\epsilon_{m}0^{-}}
\nonumber\\
&\quad\times
\mathrm{Tr}_{\rho,s}
\left[
B(\bm{k}')
\left[
\tilde{\mathcal{G}}^{\mathcal{A}}
(k_{+}',k_{-}')
\right]_{21}
\right].
\label{GeneralizedGapMinus}
\end{align}
In equilibrium,
$
\Delta_{12}^{\mathcal{A}=0}(0)
=
\Delta
$
and
$
\Delta_{21}^{\mathcal{A}=0}(0)
=
\Delta^{*}
$,
and we choose $\Delta$ to be real in this work.
Eq.~\eqref{OrbitalSelfEnergy} is generalized to four-momentum space as
\begin{align}
&
\tilde{\Sigma}^{\mathcal{A}}
(k_{+},k_{-})
\nonumber\\
&=
-
f_{c}(k^{2})
\left[
\tau_{+}\otimes B(\bm{k})\,
\Delta_{12}^{\mathcal{A}}(q)
+
\tau_{-}\otimes B^{\dagger}(\bm{k})\,
\Delta_{21}^{\mathcal{A}}(q)
\right].
\label{ExternalFieldSelfEnergy}
\end{align}
The Dyson equation in the presence of the external field is then written as
\begin{align}
\left[
\tilde{\mathcal{G}}^{\mathcal{A}}
\right]^{-1}
&=
\left[
\tilde{\mathcal{G}}_{0}^{\mathcal{A}}
\right]^{-1}
-
\tilde{\Sigma}^{\mathcal{A}}
.
\label{DysonEquationExternalField}
\end{align}
The full vertex function and the bare vertex are defined through functional derivatives of the inverse Green functions with respect to the field as
\begin{align}
\tilde{\Gamma}
_{i}(k_{1},k_{2};q)
&=
\left.
\frac{
\delta
\left[
\tilde{\mathcal{G}}^{\mathcal{A}}
\right]^{-1}
(k_{1},k_{2})
}{
\delta\mathcal{A}_{i}(q)
}
\right|_{\mathcal{A}=0},
\label{FullVertexDefinition}
\\
\tilde{\Gamma}^{(0)}
_{i}(k_{1},k_{2};q)
&=
\left.
\frac{
\delta
\left[
\tilde{\mathcal{G}}_{0}^{\mathcal{A}}
\right]^{-1}
(k_{1},k_{2})
}{
\delta\mathcal{A}_{i}(q)
}
\right|_{\mathcal{A}=0}.
\label{BareVertexDefinition}
\end{align}
Using these definitions and functionally differentiating the Dyson equation~(\ref{DysonEquationExternalField}) with respect to $\mathcal{A}_{i}(q)$, we obtain
\begin{align}
\tilde{\Gamma}
_{i}
&=
\tilde{\Gamma}^{(0)}
_{i}
-
\left.
\frac{
\delta\tilde{\Sigma}^{\mathcal{A}}
}{
\delta\mathcal{A}_{i}(q)
}
\right|_{\mathcal{A}=0}.
\label{VertexFromDysonEquation}
\end{align}
Next, differentiating
$
\tilde{\mathcal{G}}^{\mathcal{A}}
\left[
\tilde{\mathcal{G}}^{\mathcal{A}}
\right]^{-1}
=
1
$
with respect to the external field gives
\begin{align}
\frac{
\delta\tilde{\mathcal{G}}^{\mathcal{A}}
}{
\delta\mathcal{A}_{i}
}
&=
-
\tilde{\mathcal{G}}^{\mathcal{A}}
\frac{
\delta
\left[
\tilde{\mathcal{G}}^{\mathcal{A}}
\right]^{-1}
}{
\delta\mathcal{A}_{i}
}
\tilde{\mathcal{G}}^{\mathcal{A}}.
\end{align}
Setting the external field to zero gives
\begin{align}
\left.
\frac{
\delta\tilde{\mathcal{G}}^{\mathcal{A}}
(k_{1},k_{2})
}{
\delta\mathcal{A}_{i}(q)
}
\right|_{\mathcal{A}=0}
&=
-
\tilde{\mathcal{G}}(k_{1})
\tilde{\Gamma}
_{i}(k_{1},k_{2};q)
\tilde{\mathcal{G}}(k_{2}).
\label{GreenFunctionVariation}
\end{align}
Functionally differentiating Eqs.~(\ref{GeneralizedGapPlus}) and (\ref{GeneralizedGapMinus}) with respect to the external field and substituting Eq.~(\ref{GreenFunctionVariation}), we obtain
\begin{align}
&
\left.
\frac{
\delta\Delta_{12}^{\mathcal{A}}(q)
}{
\delta\mathcal{A}_{i}(q)
}
\right|_{\mathcal{A}=0}
\nonumber\\
&=
-
\frac{VT}{2}
\sum_{\bm{k}',m}
f_{c}(k'^{2})
e^{-i\epsilon_{m}0^{-}}
\nonumber\\
&\quad\times
\mathrm{Tr}_{\rho,s}
\Biggl[
B^{\dagger}(\bm{k}')
\left[
\tilde{\mathcal{G}}
(k_{+}')
\tilde{\Gamma}
_{i}
(k_{+}',k_{-}';q)
\tilde{\mathcal{G}}
(k_{-}')
\right]_{12}
\Biggr],
\label{GapVariationPlus}
\end{align}
and
\begin{align}
&
\left.
\frac{
\delta\Delta_{21}^{\mathcal{A}}(q)
}{
\delta\mathcal{A}_{i}(q)
}
\right|_{\mathcal{A}=0}
\nonumber\\
&=
-
\frac{VT}{2}
\sum_{\bm{k}',m}
f_{c}(k'^{2})
e^{-i\epsilon_{m}0^{-}}
\nonumber\\
&\quad\times
\mathrm{Tr}_{\rho,s}
\Biggl[
B(\bm{k}')
\left[
\tilde{\mathcal{G}}
(k_{+}')
\tilde{\Gamma}
_{i}
(k_{+}',k_{-}';q)
\tilde{\mathcal{G}}
(k_{-}')
\right]_{21}
\Biggr].
\label{GapVariationMinus}
\end{align}
Using Eqs.~(\ref{GapVariationPlus}) and (\ref{GapVariationMinus}) in the functional derivative of Eq.~(\ref{ExternalFieldSelfEnergy}) gives
\begin{align}
&
\left.
\frac{
\delta\tilde{\Sigma}^{\mathcal{A}}
(k_{+},k_{-})
}{
\delta\mathcal{A}_{i}(q)
}
\right|_{\mathcal{A}=0}
\nonumber\\
&=
f_{c}(k^{2})
\left[
\tau_{+}\otimes B(\bm{k})\,C_{i}^{+}(q)
+
\tau_{-}\otimes B^{\dagger}(\bm{k})\,C_{i}^{-}(q)
\right],
\label{SelfEnergyVariation}
\end{align}
where we have defined
$\delta \Delta^{\mathcal{A}}_{12}/\delta\mathcal{A}_{i}|_{\mathcal{A}=0}=-C_{i}^{+}(q)$
and
$\delta \Delta^{\mathcal{A}}_{21}/\delta\mathcal{A}_{i}|_{\mathcal{A}=0}=-C_{i}^{-}(q)$.
Finally, substituting Eq.~(\ref{SelfEnergyVariation}) into Eq.~(\ref{VertexFromDysonEquation}), we obtain the BSE in the Bloch-band basis,
\begin{align}
&
\tilde{\Gamma}
_{i}
(k_{+},k_{-};q)
=
\tilde{\Gamma}^{(0)}
_{i}
(k_{+},k_{-};q)
\nonumber\\
&\quad
-
f_{c}(k^{2})
\left[
\tau_{+}\otimes B(\bm{k})\,C_{i}^{+}(q)
+
\tau_{-}\otimes B^{\dagger}(\bm{k})\,C_{i}^{-}(q)
\right].
\label{OrbitalBetheSalpeterCompact}
\end{align}
In the static limit $i\omega_{\lambda}=0$, Eq.~\eqref{OrbitalBetheSalpeterCompact} reduces to Eq.~\eqref{BSEBeforePauliExpansion} in the main text.

We briefly discuss the extension of the above derivation to intraorbital \(s\)-wave pairing in the Bloch-band basis. An important point is that the dimensionality of the resulting BSE is determined by the rank of the pairing interaction rather than by the detailed matrix structure of the pairing. 
In general, a finite-rank separable pairing interaction can be written as
\begin{align}
H^{(r)}_{\mathrm{int}}
=-\sum_{\boldsymbol q}\sum_{\alpha=1}^{r}
V_\alpha P_{\alpha,\boldsymbol q}^{\dagger}
P_{\alpha,\boldsymbol q}.
\end{align}
Here, \(r\) denotes the rank of the interaction in the pairing-channel space, \(P_{\alpha,\boldsymbol q}\) is the pair-annihilation operator in channel \(\alpha\), and \(V_\alpha\) is the corresponding coupling constant.
A decomposition of the pairing interaction into independent channels is also employed in the collective-mode analysis of Ref.~\cite{levitan2024linear}. 
The corresponding mean-field self-energy depends on $r$ complex order parameters,
$\Delta_{\alpha,\bm{q}}
=
V_\alpha\langle 
P_{\alpha,\bm{q}}\rangle$, and their complex conjugates. Consequently, the BSE can be reduced to a $2r\times2r$ matrix equation for each $\bm{q}$.

In the present study, the interaction in Eq. (A4) 
corresponds to the rank-one case ($r=1$). Thus, the mean-field self-energy depends only on a single complex order parameter 
$\Delta_{\bm{q}}
=
V\langle P_{\bm{q}}\rangle$ 
and its complex conjugate, leading to the $2\times2$ BSE in Eq. (34). 
Similarly, for intraorbital $s$-wave pairing, a rank-one separable interaction also leads to a 
$2\times2$ BSE. 
In this case, the pairing matrix in the Bloch-band basis differs from $B(\bm{k})$ defined in Eq. (A1), which describes the intraband pairing considered here. 
This difference modifies the matrix elements of the BSE but not its dimensionality. 
If independent pairing interactions are introduced for different orbitals, the rank of the interaction may increase, resulting in a higher-dimensional BSE. 
Nevertheless, the reduction to a finite-dimensional matrix equation remains possible as long as the interaction has a finite-rank separable form.

However, reducing the BSE to a finite-dimensional matrix equation does not necessarily guarantee that the resulting physical quantities can be obtained analytically, since the matrix elements of the reduced BSE and the electromagnetic-response kernel generally involve momentum integrals. 
In the present study, we derive the required relations between the vertex correction and the electromagnetic response without explicitly evaluating these momentum integrals, by using the gap equation and total-derivative identities, as demonstrated in Appendices B and D. 
Whether a similar analytical treatment is possible for intraorbital pairing or higher-rank interactions depends on the detailed pairing structure and interaction.

%
\section{Derivation of Eqs.
(\ref{eq:chi_a_explicit})
 and 
(\ref{eq:ratio_long_wavelength})}
\label{apdxB}
Using the definitions in Sec.~V, the antisymmetric source term and kernel can be written as
\begin{align}
\chi_{a,i}(\bm q)
&=
-\frac{VT}{2\sqrt{2}}
\sum_{\bm k,n}
e^{-i\epsilon_n0^-}
\operatorname{Tr}_{N,\rho,s}
\left[
\tilde{\mathcal G}_+
\tilde\Gamma_i^{(0)}
\tilde{\mathcal G}_-X
\right],
\label{eq:appB_chi_a}
\\
\Pi_a(\bm q)
&=
\frac{VT}{4}
\sum_{\bm k,n}
e^{-i\epsilon_n0^-}
\operatorname{Tr}_{N,\rho,s}
\left[
\tilde{\mathcal G}_+X
\tilde{\mathcal G}_-X
\right].
\label{eq:appB_Pi_a}
\end{align}
Here, we have defined
$
X(\bm k)
=
f_c(k^2)\tilde V_a(\bm k)
$
and
$
X_\pm=X(\bm k_\pm)
$.

We derive the relation between $\chi_{a,i}$ and $\Pi_a$ in the long-wavelength limit.
For the bare current vertex~\eqref{BareCurrentVertex},
\begin{align}
\tau_3
\left[
\tilde{\mathcal H}_{0,\mathrm{BdG}}(\bm k_+)
-
\tilde{\mathcal H}_{0,\mathrm{BdG}}(\bm k_-)
\right]
=
2\sum_{i=1}^{3}q_i\tilde\Gamma_i^{(0)}
\label{eq:B11}
\end{align}
holds.
Substituting
$
\tilde{\mathcal H}_{\mathrm{BdG}}(\bm k)
=
\tilde{\mathcal H}_{0,\mathrm{BdG}}(\bm k)
-
\Delta\tau_3X(\bm k)
$ 
into Eq.~\eqref{eq:B11} gives
\begin{align}
2\sum_{i=1}^{3}q_i\tilde\Gamma_i^{(0)}
=
-\tau_3
\left(
\tilde{\mathcal G}_+^{-1}
-
\tilde{\mathcal G}_-^{-1}
\right)
+
\Delta(X_+-X_-).
\label{eq:appB_bare_vertex}
\end{align}
Using
$\{\tau_3,X\}=0$
and
$[\tau_3,\tilde{\mathcal H}_{0,\mathrm{BdG}}]=0$,
we find
\begin{align}
\left[
\tau_3,
\tilde{\mathcal G}_\pm^{-1}
\right]
=
2\Delta X_\pm.
\label{eq:appB_commutator}
\end{align}
Multiplying Eq.~\eqref{eq:appB_bare_vertex} by $\tilde{\mathcal G}_+$ from the left and by $\tilde{\mathcal G}_-$ from the right, and using Eq.~\eqref{eq:appB_commutator}, we obtain
\begin{align}
2\tilde{\mathcal G}_+
\left(
\sum_{i=1}^{3}q_i\tilde\Gamma_i^{(0)}
\right)
\tilde{\mathcal G}_-
&=
\tilde{\mathcal G}_+\tau_3
-
\tau_3\tilde{\mathcal G}_-
\nonumber\\
&\quad-
\Delta
\tilde{\mathcal G}_+
(X_++X_-)
\tilde{\mathcal G}_-.
\label{eq:appB_contracted_vertex}
\end{align}
At $\bm q=\bm 0$, Eq.~\eqref{eq:appB_commutator} gives
\begin{align}
\tilde{\mathcal G}\tau_3
-
\tau_3\tilde{\mathcal G}
=
2\Delta
\tilde{\mathcal G}X\tilde{\mathcal G},
\label{eq:appB_G_tau3}
\end{align}
where 
$
\tilde{\mathcal G}
=
\tilde{\mathcal G}(\bm{k},i\epsilon_{n})
$.
The gap equation also gives
\begin{align}
\Pi_a(\bm 0)
=
\frac{VT}{4}
\sum_{\bm k,n}
e^{-i\epsilon_n0^-}
\operatorname{Tr}_{N,\rho,s}
\left[
\tilde{\mathcal G}X
\tilde{\mathcal G}X
\right]
=
1.
\label{eq:appB_gap}
\end{align}
Contracting Eq.~\eqref{eq:appB_chi_a} with $\bm q$ and using Eqs.~\eqref{eq:appB_contracted_vertex}, \eqref{eq:appB_G_tau3}, and \eqref{eq:appB_gap}, we obtain
\begin{align}
\sum_{i=1}^{3}q_i\chi_{a,i}(\bm q)
=
-\sqrt{2}\Delta
\left[
1-\Pi_a(\bm q)
\right]
+
\mathcal R(\bm q),
\label{eq:appB_qchi_remainder}
\end{align}
where we have defined $\mathcal R(\bm q)$ as
\begin{align}
\mathcal R(\bm q)
&=
\frac{VT\Delta}{4\sqrt{2}}
\sum_{\bm k,n}
e^{-i\epsilon_n0^-}
\nonumber\\
&\times
\operatorname{Tr}_{N,\rho,s}
\left[
\tilde{\mathcal G}_+
(X_++X_--2X)
\tilde{\mathcal G}_-X
\right]
\nonumber\\
&
-
\frac{VT}{4\sqrt{2}}
\sum_{\bm k,n}
e^{-i\epsilon_n0^-}
\nonumber\\
&\times
\operatorname{Tr}_{N,\rho,s}
\left[
\left\{
\tilde{\mathcal G}_+\tau_3
-
\tau_3\tilde{\mathcal G}_-
-
\tilde{\mathcal G}\tau_3
+
\tau_3\tilde{\mathcal G}
\right\}X
\right].
\label{eq:appB_remainder}
\end{align}
We now show that $\mathcal R(\bm q)$ vanishes up to
second order in $q$.
Using the inversion operator, we have
\begin{align}
\tilde{\mathcal G}(-\bm k,i\epsilon_n)
&=
\mathcal{I}
\tilde{\mathcal G}(\bm k,i\epsilon_n)
\mathcal{I}^{-1},
\\
X(-\bm k)
&=
\mathcal{I}
X(\bm k)
\mathcal{I}^{-1}.
\end{align}
Under $\bm q\rightarrow-\bm q$, we have
$\tilde{\mathcal G}_+\leftrightarrow\tilde{\mathcal G}_-$
and
$X_+\leftrightarrow X_-$.
After changing the integration variable as $\bm k\rightarrow-\bm k$ in the momentum sum, Eq.~\eqref{eq:appB_remainder} returns to its original form.
Therefore,
\begin{align}
\mathcal R(-\bm q)
=
\mathcal R(\bm q).
\label{eq:appB_R_even}
\end{align}
Thus, only even-order terms appear in the long-wavelength expansion of $\mathcal R(\bm q)$.
In particular, using
\begin{align}
X_++X_--2X
=
\frac{1}{4}
\sum_{\ell=1}^{3}
\sum_{m=1}^{3}
q_\ell q_m
\partial_\ell\partial_mX
+
O(q^4)
\end{align}
and
\begin{align}
&\tilde{\mathcal G}_+\tau_3
-
\tau_3\tilde{\mathcal G}_-
-
\tilde{\mathcal G}\tau_3
+
\tau_3\tilde{\mathcal G}
\nonumber\\
&=
\frac{1}{2}
\sum_{\ell=1}^{3}
q_\ell
\left[
(\partial_\ell\tilde{\mathcal G})\tau_3
+
\tau_3(\partial_\ell\tilde{\mathcal G})
\right]
\nonumber\\
&\quad+
\frac{1}{8}
\sum_{\ell=1}^{3}
\sum_{m=1}^{3}
q_\ell q_m
\left[
\partial_\ell\partial_m\tilde{\mathcal G},
\tau_3
\right]
+
O(q^3),
\end{align}
we obtain
\begin{align}
\sum_{i=1}^{3}q_i\chi_{a,i}(\bm q)
=
-\sqrt{2}\Delta
\left[
1-\Pi_a(\bm q)
\right]
+
\mathcal R^{(2)}(\bm q)
+
O(q^4),
\label{eq:appB_qchi_q2}
\end{align}
where
\begin{align}
\mathcal R^{(2)}(\bm q)
&=
\frac{VT\Delta}{16\sqrt{2}}
\sum_{\ell=1}^{3}
\sum_{m=1}^{3}
q_\ell q_m
\sum_{\bm k,n}
e^{-i\epsilon_n0^-}
\nonumber\\
&\times
\operatorname{Tr}_{N,\rho,s}
\left[
\tilde{\mathcal G}
(\partial_\ell\partial_mX)
\tilde{\mathcal G}X
\right]
\nonumber\\
&\quad-
\frac{VT}{32\sqrt{2}}
\sum_{\ell=1}^{3}
\sum_{m=1}^{3}
q_\ell q_m
\sum_{\bm k,n}
e^{-i\epsilon_n0^-}
\nonumber\\
&\times
\operatorname{Tr}_{N,\rho,s}
\left[
\left[
\partial_\ell\partial_m\tilde{\mathcal G},
\tau_3
\right]X
\right].
\label{eq:appB_R2}
\end{align}
Differentiating Eq.~\eqref{eq:appB_G_tau3} twice with respect to momentum gives
\begin{align}
\left[
\partial_\ell\partial_m\tilde{\mathcal G},
\tau_3
\right]
=
2\Delta
\partial_\ell\partial_m
\left(
\tilde{\mathcal G}X\tilde{\mathcal G}
\right).
\label{eq:appB_second_derivative}
\end{align}
Substituting this into Eq.~\eqref{eq:appB_R2} and using the cyclic property of the trace and the product rule, we find that terms antisymmetric under $\ell\leftrightarrow m$ vanish because $q_\ell q_m$ is symmetric under $\ell\leftrightarrow m$.
The remaining terms give
\begin{align}
\mathcal R^{(2)}(\bm q)
&=
\frac{VT\Delta}{16\sqrt{2}}
\sum_{\ell=1}^{3}
\sum_{m=1}^{3}
q_\ell q_m
\sum_{\bm k,n}
e^{-i\epsilon_n0^-}
\nonumber\\
&
\times
\partial_\ell
\operatorname{Tr}_{N,\rho,s}
\Biggl[
\tilde{\mathcal G}X\tilde{\mathcal G}
(\partial_mX)
-
\partial_m
\left(
\tilde{\mathcal G}X\tilde{\mathcal G}
\right)
X
\Biggr].
\label{eq:appB_R2_total_derivative}
\end{align}
In the thermodynamic limit, replacing the momentum sum by an integral gives
\begin{align}
\mathcal R^{(2)}(\bm q)
&=
\frac{VT\Delta}{16\sqrt{2}}
\sum_{\ell=1}^{3}
\sum_{m=1}^{3}
q_\ell q_m
\sum_n
e^{-i\epsilon_n0^-}
\int
\frac{d^3k}{(2\pi)^3}
\nonumber\\
&
\times
\partial_\ell
\operatorname{Tr}_{N,\rho,s}
\Biggl[
\tilde{\mathcal G}X\tilde{\mathcal G}
(\partial_mX)
-
\partial_m
\left(
\tilde{\mathcal G}X\tilde{\mathcal G}
\right)
X
\Biggr].
\end{align}
The above integral is determined by the boundary terms at $k_\ell=\pm\infty$.
Since $X$ and $\partial_mX$ vanish at these boundaries, we obtain
\begin{align}
\mathcal R^{(2)}(\bm q)
=
0.
\label{eq:appB_R2_zero}
\end{align}
With $\mathcal R^{(2)}(\bm q)=0$, Eq.~\eqref{eq:appB_qchi_q2} reduces to
\begin{align}
\sum_{i=1}^{3}q_i\chi_{a,i}(\bm q)
=
-\sqrt{2}\Delta
\left[
1-\Pi_a(\bm q)
\right]
+
O(q^4).
\label{eq:appB_qchi_final}
\end{align}

Finally, we evaluate $\chi_{a,i}(\bm q)$ directly and obtain its expression.
Expanding the Green functions in $\bm q$, we find
\begin{align}
\chi_{a,i}(\bm{q})
&=
-
\frac{V}{4\sqrt{2}}
\sum_{\ell=1}^{3}
q_{\ell}
\sum_{\bm{k}}
f_{c}(k^{2})
T\sum_{n}
e^{-i\epsilon_{n}0^{-}}
\nonumber\\
&\quad\times
\mathrm{Tr}_{N,\rho,s}
\Bigl[
\left(
\partial_{\ell}\tilde{\mathcal{G}}
\right)
\tilde{\Gamma}_{i}^{(0)}
\tilde{\mathcal{G}}
\tilde{V}_{a}(\bm{k})
\nonumber\\
&\qquad\qquad
-
\tilde{\mathcal{G}}
\tilde{\Gamma}_{i}^{(0)}
\left(
\partial_{\ell}\tilde{\mathcal{G}}
\right)
\tilde{V}_{a}(\bm{k})
\Bigr]
+
O(q^{3}).
\label{eq:chi_a_linear_trace}
\end{align}
Carrying out the trace over the internal degrees of freedom and the Matsubara-frequency sum, we obtain
\begin{align}
&
\chi_{a,i}(\bm{q})
\nonumber\\
&=
-
\frac{Vv^{2}\Delta}{12\sqrt{2}}
q_{i}
\sum_{\bm{k}}
\Biggl[
\frac{
v^{2}k^{2}f_{c}^{2}(k^{2})
}{
\epsilon_{\bm{k}}^{2}
}
\left(
\frac{1}{E_{-}^{3}(\bm{k})}
+
\frac{1}{E_{+}^{3}(\bm{k})}
\right)
\nonumber\\
&\quad
+
\frac{
2k^{2}f_{c}(k^{2})
}{
\epsilon_{\bm{k}}
}
\left.
\frac{df_{c}(s)}{ds}
\right|_{s=k^{2}}
\left(
\frac{\xi_{-}(\bm{k})}{E_{-}^{3}(\bm{k})}
-
\frac{\xi_{+}(\bm{k})}{E_{+}^{3}(\bm{k})}
\right)
\Biggr]
\nonumber\\
&
\quad
+
O(q^{3}).
\label{eq:chi_a_direct_result}
\end{align}
Using the definition of $\mathcal{S}_{1}(\bm{k})$ in Eq.~\eqref{eq:def_S1} and
$S_{1}=\sum_{\bm{k}}\mathcal{S}_{1}(\bm{k})$,
the antisymmetric source term becomes
\begin{align}
\chi_{a,i}(\bm{q})
&=
-\frac{Vv^{2}\Delta}{12\sqrt{2}}
S_{1}q_{i}
+
O(q^{3}).
\label{eq:chi_a_explicit_apdx}
\end{align}
This reproduces Eq.~\eqref{eq:chi_a_explicit} in the main text.
Contracting this result with $q_i$ gives
\begin{align}
\sum_{i=1}^{3}
q_{i}\chi_{a,i}(\bm{q})
&=
-\frac{Vv^{2}\Delta}{12\sqrt{2}}
S_{1}q^{2}
+
O(q^{4}).
\label{eq:qchi_a_explicit_apdx}
\end{align}
Comparing Eq.~(\ref{eq:qchi_a_explicit_apdx}) with Eq.~(\ref{eq:appB_qchi_final}), we obtain
\begin{align}
1-\Pi_{a}(\bm{q})
&=
\frac{Vv^{2}}{24}
S_{1}q^{2}
+
O(q^{4}).
\label{eq:Pi_a_from_chi}
\end{align}
Combining Eqs.~(\ref{eq:chi_a_explicit_apdx}) and (\ref{eq:Pi_a_from_chi}) reproduces Eq.~(\ref{eq:ratio_long_wavelength}) in the main text.

\section{Properties of the symmetric pairing channel}
\label{apdxC}
\subsection{Absence of a gapless mode in the symmetric channel}

In this Appendix, we show that
$1-\Pi_s(\bm{0})\neq0$
for the symmetric pairing channel.
Using Eq.~\eqref{BSEKernelDefinition} and the definitions of
$\Pi_s$ and $\Pi_a$ in the main text, the kernels in the two channels at $\bm q=0$ can be written as
\begin{align}
\Pi_s(\bm{0})
&=
-\frac{VT}{4}
\sum_{\bm{k},n}
f_c^2(k^2)e^{-i\epsilon_n0^-}
\mathrm{Tr}_{N,\rho,s}
\left[
\tilde{\mathcal{G}} \tilde{V}_s
\tilde{\mathcal{G}} \tilde{V}_s
\right],
\label{eq:C1}
\\
\Pi_a(\bm{0})
&=
\frac{VT}{4}
\sum_{\bm{k},n}
f_c^2(k^2)e^{-i\epsilon_n0^-}
\mathrm{Tr}_{N,\rho,s}
\left[
\tilde{\mathcal{G}} \tilde{V}_a
\tilde{\mathcal{G}} \tilde{V}_a
\right].
\label{eq:C2}
\end{align}
Here, we define the symmetric pairing vertex as
\begin{align}
\tilde{V}_s(\bm{k})
=
\tau_+\otimes B(\bm{k})
+
\tau_-\otimes B^\dagger(\bm{k}).
\label{eq:Vs_app}
\end{align}
The definition of $\tilde{V}_a(\bm{k})$ is given in Eq.~\eqref{AntisymmetricPairVertex}.
A straightforward calculation gives
$
\tilde{V}_a^2
=
-1
$
and
$
\tilde{V}_a\tilde{\mathcal{G}}
=
-\tilde{\mathcal{G}}^\dagger \tilde{V}_a
$.
It then follows that
\begin{align}
\mathrm{Tr}_{N,\rho,s}
\left[
\tilde{\mathcal{G}} \tilde{V}_a
\tilde{\mathcal{G}} \tilde{V}_a
\right]
=
\mathrm{Tr}_{N,\rho,s}
\left[
\tilde{\mathcal{G}}\tilde{\mathcal{G}}^\dagger
\right].
\label{eq:C4}
\end{align}

For the symmetric vertex, the relations
$
\tilde{V}_s
=
\tau_3\tilde{V}_a
$
and
$
\tilde{V}_a\tau_3\tilde{V}_a
=
\tau_3
$
give
\begin{align}
-\mathrm{Tr}_{N,\rho,s}
\left[
\tilde{\mathcal{G}} \tilde{V}_s
\tilde{\mathcal{G}} \tilde{V}_s
\right]
=
\mathrm{Tr}_{N,\rho,s}
\left[
\tilde{\mathcal{G}}\tau_3
\tilde{\mathcal{G}}^\dagger\tau_3
\right].
\label{eq:C7}
\end{align}
Taking the difference between Eqs.~\eqref{eq:C1} and \eqref{eq:C2} and using Eqs.~\eqref{eq:C4} and \eqref{eq:C7}, we obtain
\begin{align}
\Pi_a(\bm{0})-\Pi_s(\bm{0})
&=
\frac{VT}{4}
\sum_{\bm{k},n}
f_c^2(k^2)e^{-i\epsilon_n0^-}
\nonumber\\
&\times
\left\{
\mathrm{Tr}_{N,\rho,s}
\left[
\tilde{\mathcal{G}}\tilde{\mathcal{G}}^\dagger
\right]
-
\mathrm{Tr}_{N,\rho,s}
\left[
\tilde{\mathcal{G}}\tau_3\tilde{\mathcal{G}}^\dagger\tau_3
\right]
\right\}.
\label{eq:diff_PIaPIs}
\end{align}
The following relation also holds:
\begin{align}
&
\mathrm{Tr}_{N,\rho,s}
\left[
\tilde{\mathcal{G}}\tilde{\mathcal{G}}^\dagger
\right]
-
\mathrm{Tr}_{N,\rho,s}
\left[
\tilde{\mathcal{G}}\tau_3\tilde{\mathcal{G}}^\dagger\tau_3
\right]
\nonumber\\
&=
\frac{1}{2}
\mathrm{Tr}_{N,\rho,s}
\left[
\left(
\tilde{\mathcal{G}}-\tau_3\tilde{\mathcal{G}}\tau_3
\right)
\left(
\tilde{\mathcal{G}}-\tau_3\tilde{\mathcal{G}}\tau_3
\right)^\dagger
\right]
\geq0.
\end{align}
In the last inequality, we have used the nonnegativity of the Hilbert--Schmidt norm,
$\sqrt{\mathrm{Tr}(AA^{\dagger})}\geq0$
for an arbitrary matrix $A$.
Moreover,
\begin{align}
\tilde{\mathcal{G}}-\tau_3\tilde{\mathcal{G}}\tau_3
=
-2\Delta f_c(k^2)
\tau_3\tilde{\mathcal{G}} \tilde{V}_a\tilde{\mathcal{G}}.
\end{align}
In the superconducting state, $\Delta\neq0$, and
$\tilde{\mathcal{G}}$, $\tau_3$, and $\tilde{V}_a$ are all invertible.
These properties imply that, in the region where $f_c(k^2)\neq0$,
$
\tilde{\mathcal{G}}-\tau_3\tilde{\mathcal{G}}\tau_3
\neq0
$.
It follows that
\begin{align}
\mathrm{Tr}_{N,\rho,s}
\left[
\left(
\tilde{\mathcal{G}}-\tau_3\tilde{\mathcal{G}}\tau_3
\right)
\left(
\tilde{\mathcal{G}}-\tau_3\tilde{\mathcal{G}}\tau_3
\right)^\dagger
\right]
>0.
\end{align}
Hence,
\begin{align}
\Pi_a(\bm{0})-\Pi_s(\bm{0})>0.
\label{eq:C10}
\end{align}
Since the Matsubara-frequency sum in Eq.~\eqref{eq:diff_PIaPIs} is convergent, the convergence factor
$e^{-i\epsilon_{n} 0^{-}}$
has been set to unity in the above argument.
Finally, combining 
Eq.~\eqref{eq:C10} 
with
$
\Pi_a(\bm{0})=1,
$
which follows from the gap equation, gives
\begin{align}
1-\Pi_s(\bm{0})>0.
\end{align}
In particular, this proves that
$
1-\Pi_s(\bm{0})\neq0
$.
\subsection{Vanishing of the symmetric-channel contribution in the long-wavelength limit}
The symmetric source term can be written as
\begin{align}
\chi_{s,i}(\bm{q})
&=
\frac{VT}{2\sqrt{2}}
\sum_{\bm{k},n}
f_c(k^2)e^{-i\epsilon_n0^-}
\nonumber\\
&
\times
\mathrm{Tr}_{N,\rho,s}
\left[
\tilde{\mathcal{G}}_-\tilde{V}_s(\bm{k})
\tilde{\mathcal{G}}_+\tilde{\Gamma}_i^{(0)}
\right].
\label{eq:chis_trace_app}
\end{align}
We first consider $\bm{q}=\bm{0}$.
In this case, $\tilde{\mathcal{G}}_+=\tilde{\mathcal{G}}_-=\tilde{\mathcal{G}}$, and hence
\begin{align}
\chi_{s,i}(\bm{0})
&=
\frac{VT}{2\sqrt{2}}
\sum_{\bm{k},n}
f_c(k^2)e^{-i\epsilon_n0^-}
\mathrm{Tr}_{N,\rho,s}
\left[
\tilde{\mathcal{G}}\tilde{V}_s(\bm{k})
\tilde{\mathcal{G}}\tilde{\Gamma}_i^{(0)}
\right].
\label{eq:chis_q0_app}
\end{align}
Using the inversion operator $\mathcal{I}$, we have
\begin{align}
\tilde{V}_s(-\bm{k})
=
\mathcal{I}
\tilde{V}_s(\bm{k})
\mathcal{I}^{-1},
\label{eq:Vs_minus_app}
\end{align}
and
\begin{align}
\tilde{\Gamma}_i^{(0)}
=
-\mathcal{I}\tilde{\Gamma}_i^{(0)}\mathcal{I}^{-1}.
\label{eq:Gamma_minus_app}
\end{align}

Using Eqs.~\eqref{eq:Vs_minus_app} and \eqref{eq:Gamma_minus_app}, the integrand in Eq.~\eqref{eq:chis_q0_app} transforms under $\bm{k}\to-\bm{k}$ as
\begin{align}
&
\mathrm{Tr}_{N,\rho,s}
\left[
\tilde{\mathcal{G}}(-\bm{k},i\epsilon_{n})
\tilde{V}_s(-\bm{k})
\tilde{\mathcal{G}}(-\bm{k},i\epsilon_{n})
\tilde{\Gamma}_i^{(0)}
\right]
\nonumber\\
&=
\mathrm{Tr}_{N,\rho,s}
\Bigl[
\tilde{\mathcal{G}}(\bm{k},i\epsilon_{n})
\tilde{V}_s(\bm{k})
\tilde{\mathcal{G}}(\bm{k},i\epsilon_{n})
\mathcal{I}
\tilde{\Gamma}_i^{(0)}
\mathcal{I}^{-1}
\Bigr]
\nonumber\\
&=
-
\mathrm{Tr}_{N,\rho,s}
\left[
\tilde{\mathcal{G}}(\bm{k},i\epsilon_{n})
\tilde{V}_s(\bm{k})
\tilde{\mathcal{G}}(\bm{k},i\epsilon_{n})
\tilde{\Gamma}_i^{(0)}
\right].
\label{eq:integrand_odd_s}
\end{align}
Equation~\eqref{eq:integrand_odd_s} shows that the integrand is an odd function of $\bm{k}$.
It follows that
\begin{align}
\chi_{s,i}(\bm{0})=0.
\label{eq:chis_zero_app}
\end{align}
Finally, from Eq.~\eqref{SymmetricBSE} in the main text,
\begin{align}
C_{s,i}(\bm{q})
=
\frac{\chi_{s,i}(\bm{q})}
{1-\Pi_s(\bm{q})}.
\end{align}
Using
$1-\Pi_s(\bm{0})\neq 0$,
as shown in the preceding section, we obtain
\begin{align}
\lim_{\bm{q}\to 0}
C_{s,i}(\bm{q})
=
\frac{\chi_{s,i}(\bm{0})}{1-\Pi_s(\bm{0})}
=
0.
\label{eq:Cs_order}
\end{align}
The vanishing of $C_{s,i}$ shows that the symmetric pairing channel does not contribute to the electromagnetic response in the long-wavelength limit.
\section{Derivation of Eq.~(\ref{SubtractedLongitudinalResponseExplicit})}
\label{apdxD}
Using the results of Mizoguchi and Ogata~\cite{mizoguchi2015meissner}, we evaluate the bare responses in the superconducting and normal states at zero temperature.
The bare electromagnetic response in the superconducting state is given by
\begin{align}
\Phi_{xx}^{(0),\mathrm{SC}}(\bm{0})
&=
-2e^2v^2
\sum_{\bm{k}}
\left(
1-\frac{v^2k_x^2}{\epsilon_{\bm{k}}^2}
\right)
\nonumber\\
&\times
\Biggl[
\left(
1+
\frac{
\xi_+(\bm{k})\xi_-(\bm{k})-\Delta_{\bm{k}}^2
}{
E_+(\bm{k})E_-(\bm{k})
}
\right)
\nonumber\\
&\times
\frac{
f(E_-(\bm{k}))-f(E_+(\bm{k}))
}{
E_+(\bm{k})-E_-(\bm{k})
}
\nonumber\\
&
-
\left(
1-
\frac{
\xi_+(\bm{k})\xi_-(\bm{k})-\Delta_{\bm{k}}^2
}{
E_+(\bm{k})E_-(\bm{k})
}
\right)
\nonumber\\
&
\times
\frac{
f(E_-(\bm{k}))+f(E_+(\bm{k}))-1
}{
E_+(\bm{k})+E_-(\bm{k})
}
\Biggr].
\label{eq:D1}
\end{align}
The above expression contains only the interband contribution of Ref.~\cite{mizoguchi2015meissner}, because the intraband contribution vanishes at zero temperature and is omitted here.
Furthermore, by isotropy, all diagonal components of the electromagnetic response kernel are equal at $\bm{q}=\bm{0}$, and their common value gives the $O(|\bm{q}|^{0})$ term of the longitudinal response.
We take $\bm{q}$ along the $x$ direction, for which $\Phi_{xx}$ gives the longitudinal response to the electromagnetic field.

For Eq.~\eqref{eq:D1}, 
taking $T=0$ and performing the angular average
$
k_i k_j
\to
k^2\delta_{ij}/3
$,
we obtain
\begin{align}
\Phi_{xx}^{(0),\mathrm{SC}}(\bm{0})
&=
-2e^2
\sum_{\bm{k}}
\frac{
v^2(2\epsilon_{\bm{k}}^2+M^2)
}{
6\epsilon_{\bm{k}}^3
}
\left(
\frac{\xi_{+}(\bm{k})}{E_{+}(\bm{k})}
-
\frac{\xi_{-}(\bm{k})}{E_{-}(\bm{k})}
\right).
\end{align}
We define the integrand as
\begin{align}
\mathcal{S}_2(\bm{k})
=
-\frac{
v^2(2\epsilon_{\bm{k}}^2+M^2)
}{
6\epsilon_{\bm{k}}^3
}
\left(
\frac{\xi_{+}(\bm{k})}{E_{+}(\bm{k})}
-
\frac{\xi_{-}(\bm{k})}{E_{-}(\bm{k})}
\right).
\label{eq:S4_simple}
\end{align}
We also define
$
S_2
\equiv
\sum_{\bm{k}}\mathcal{S}_2(\bm{k})
$.
Thus, 
\begin{align}
\lim_{\bm{q}\to 0}
\Phi_{\mathrm{L}}^{(0),\mathrm{SC}}(\bm{q})
=
2e^2S_2.
\end{align}

In the normal state, both intraband and interband contributions are present and are given, respectively, by
\begin{align}
\Phi_{xx}^{\mathrm{intra},\mathrm{N}}(\bm{0})
&=
2e^2v^2
\sum_{\bm{k}}
\frac{v^2k_x^2}{\epsilon_{\bm{k}}^2}
\left[
\frac{\partial f(\xi_+(\bm{k}))}{\partial \xi_+(\bm{k})}
+
\frac{\partial f(\xi_-(\bm{k}))}{\partial \xi_-(\bm{k})}
\right],
\label{eq:normal_intra_start}
\\
\Phi_{xx}^{\mathrm{inter},\mathrm{N}}(\bm{0})
&=
-2e^2v^2
\sum_{\bm{k}}
\left(
1-\frac{v^2k_x^2}{\epsilon_{\bm{k}}^2}
\right)
\nonumber\\
&\times
\frac{
f(\xi_-(\bm{k}))-f(\xi_+(\bm{k}))
}{
\epsilon_{\bm{k}}
}.
\label{eq:normal_inter_start}
\end{align}
Here, $f(\xi)$ denotes the Fermi distribution function.
At zero temperature, combining these two contributions gives the following kernel:
\begin{align}
\mathcal{S}_3(\bm{k})
&=
\frac{
v^2(2\epsilon_{\bm{k}}^2+M^2)
}{
3\epsilon_{\bm{k}}^3
}
\left[
f(\xi_{-}(\bm{k}))-f(\xi_{+}(\bm{k}))
\right]
\nonumber\\
&\quad+
\frac{
v^2(\epsilon_{\bm{k}}^2-M^2)
}{
3\epsilon_{\bm{k}}^2
}
\left[
\delta(\xi_{-}(\bm{k}))
+
\delta(\xi_{+}(\bm{k}))
\right].
\label{eq:S5_definition}
\end{align}
Defining
$
S_3
\equiv
\sum_{\bm{k}}\mathcal{S}_3
(\bm{k})
$,
the bare longitudinal response in the normal state becomes
\begin{align}
\lim_{\bm{q}\to 0}
\Phi_{\mathrm{L}}^{(0),\mathrm{N}}(\bm{q})
=
-2e^2S_3.
\end{align}
Subtracting the normal-state response from the superconducting-state response gives
\begin{align}
\lim_{\bm{q}\to 0}
\Phi_{\mathrm{L}}^{\mathrm{sub}}(\bm{q})
=
2e^2(S_2+S_3)
.
\label{eq:Phi_sub_L_scalar}
\end{align}
We now derive the relation between $S_2+S_3$ and $S_1$.
A straightforward calculation gives
\begin{align}
\frac{d}{d|\bm{k}|}
\left(
\frac{\xi_{+}(\bm{k})}{E_{+}(\bm{k})}
-
\frac{\xi_{-}(\bm{k})}{E_{-}(\bm{k})}
\right)
=
\frac{\Delta^2\epsilon_{\bm{k}}}{|\bm{k}|}
\mathcal{S}_1(\bm{k})
\label{eq:difference_derivative_S1}
\end{align}
where $\mathcal{S}_1(\bm{k})$ is the kernel defined in Eq.~\eqref{eq:def_S1} in the main text.
We then obtain
\begin{align}
&\mathcal{S}_2(\bm{k})
-
\frac{\Delta^2v^2}{6}\mathcal{S}_1(\bm{k})
\nonumber\\
&\qquad=
-\frac{1}{k^2}
\frac{d}{d|\bm{k}|}
\left[
\frac{v^2k^3}{6\epsilon_{\bm{k}}}
\left(
\frac{\xi_{+}(\bm{k})}{E_{+}(\bm{k})}
-
\frac{\xi_{-}(\bm{k})}{E_{-}(\bm{k})}
\right)
\right].
\label{eq:S4_total_derivative}
\end{align}
Next, we derive a total-derivative expression for $\mathcal{S}_3(\bm{k})$.
At zero temperature,
\begin{align}
\frac{d}{d|\bm{k}|}
\left[
f(\xi_{-}(\bm{k}))-f(\xi_{+}(\bm{k}))
\right]
&=
\frac{v^2k}{\epsilon_{\bm{k}}}
\left[
\delta(\xi_{-}(\bm{k}))
+
\delta(\xi_{+}(\bm{k}))
\right],
\label{eq:fermi_derivative}
\end{align}
and hence
\begin{align}
\mathcal{S}_3(\bm{k})
&=
\frac{1}{k^2}
\frac{d}{d|\bm{k}|}
\left[
\frac{v^2k^3}{3\epsilon_{\bm{k}}}
\left\{
f(\xi_{-}(\bm{k}))-f(\xi_{+}(\bm{k}))
\right\}
\right].
\label{eq:S5_total_derivative}
\end{align}
Adding Eqs.~\eqref{eq:S4_total_derivative} and \eqref{eq:S5_total_derivative}, we obtain
\begin{align}
&\mathcal{S}_2(\bm{k})+\mathcal{S}_3(\bm{k})
-\frac{\Delta^2v^2}{6}\mathcal{S}_1(\bm{k})
\nonumber\\
&=
\frac{1}{k^2}
\frac{d}{d|\bm{k}|}
\Biggl[
-\frac{v^2k^3}{6\epsilon_{\bm{k}}}
\left(
\frac{\xi_{+}(\bm{k})}{E_{+}(\bm{k})}
-
\frac{\xi_{-}(\bm{k})}{E_{-}(\bm{k})}
\right)
\nonumber\\
&\hspace{17mm}
+
\frac{v^2k^3}{3\epsilon_{\bm{k}}}
\left\{
f(\xi_{-}(\bm{k}))-f(\xi_{+}(\bm{k}))
\right\}
\Biggr].
\label{eq:S4S5_total_derivative}
\end{align}
Integrating over the radial momentum gives
\begin{align}
&S_2+S_3
-\frac{\Delta^2v^2}{6}S_1
\nonumber\\
&=
\frac{1}{2\pi^2}
\Biggl[
-\frac{v^2k^3}{6\epsilon_{\bm{k}}}
\left(
\frac{\xi_{+}(\bm{k})}{E_{+}(\bm{k})}
-
\frac{\xi_{-}(\bm{k})}{E_{-}(\bm{k})}
\right)
\nonumber\\
&\hspace{17mm}
+
\frac{v^2k^3}{3\epsilon_{\bm{k}}}
\left\{
f(\xi_{-}(\bm{k}))-f(\xi_{+}(\bm{k}))
\right\}
\Biggr]_{0}^{\infty}.
\label{eq:S4S5_boundary}
\end{align}
As $|\bm{k}|\to0$, both terms inside the square brackets are proportional to $k^3$, and hence the boundary contribution at the lower limit vanishes.
At the upper limit, as $|\bm{k}|\to\infty$, for sufficiently large $|\bm{k}|$,
$\xi_{+}(\bm{k})>0$,
$\xi_{-}(\bm{k})<0$,
and
$f(\xi_{-}(\bm{k}))-f(\xi_{+}(\bm{k}))=1$.
Denoting the expression inside the square brackets in Eq.~\eqref{eq:S4S5_boundary} by $W(\bm{k})$, we can write
\begin{align}
W(\bm{k})
&=
\frac{v^2k^3\Delta^2f_c^2(k^2)}{6\epsilon_{\bm{k}}}
\Biggl[
\frac{1}{
E_{+}(\bm{k})
\left(
E_{+}(\bm{k})+\xi_{+}(\bm{k})
\right)
}
\nonumber\\
&\qquad\qquad\qquad
+
\frac{1}{
E_{-}(\bm{k})
\left(
E_{-}(\bm{k})+|\xi_{-}(\bm{k})|
\right)
}
\Biggr].
\end{align}
For sufficiently large $|\bm{k}|$, we can take
$\epsilon_{\bm{k}}\ge 2|\mu|$.
Then,
$
\xi_{+}(\bm{k})
\ge
\epsilon_{\bm{k}}-|\mu|
\ge
\epsilon_{\bm{k}}/2
$
and
$
|\xi_{-}(\bm{k})|
\ge
\epsilon_{\bm{k}}-|\mu|
\ge
\epsilon_{\bm{k}}/2
$.
Since
$E_{+}({\bm{k}})\ge\xi_{+}({\bm{k}})$
and
$E_{-}({\bm{k}})\ge|\xi_{-}({\bm{k}})|$,
it follows that
$
E_{+}({\bm{k}})
\left(
E_{+}({\bm{k}})+\xi_{+}({\bm{k}})
\right)
\ge
\epsilon_{\bm{k}}^2/2
$
and
$
E_{-}(\bm{k})
\left(
E_{-}(\bm{k})+|\xi_{-}(\bm{k})|
\right)
\ge
\epsilon_{\bm{k}}^2/2
$.
These inequalities give
\begin{align}
0
\le
W(\bm{k})
\le
\frac{2v^2\Delta^2}{3}
\frac{k^3}{\epsilon_{\bm{k}}^3}
f_c^2(k^2)
\le
\frac{2\Delta^2}{3v}
f_c^2(k^2).
\end{align}
Since
$f_c(k^2)\to0$
as
$|\bm{k}|\to\infty$,
we obtain
\begin{align}
\lim_{|\bm{k}|\to\infty}W(\bm{k})=0.
\end{align}
Since the boundary contributions at both limits vanish, Eq.~\eqref{eq:S4S5_boundary} gives
\begin{align}
S_2+S_3
=
\frac{\Delta^2v^2}{6}S_1
.
\label{eq:S4S5_S1}
\end{align}
Substituting Eq.~\eqref{eq:S4S5_S1} into Eq.~\eqref{eq:Phi_sub_L_scalar} gives Eq.~\eqref{SubtractedLongitudinalResponseExplicit} in the main text.

\bibliography{apsbib}
\end{document}